\documentclass[11pt]{article}

\usepackage{multirow} 
\usepackage{arydshln} 
\usepackage{booktabs} 
\usepackage{amsmath}
\usepackage{makecell}
\usepackage{soul, color, xcolor}
\usepackage[table]{xcolor}
\usepackage{enumitem}
\usepackage[ruled,vlined]{algorithm2e}
\usepackage[dvipsnames]{xcolor}
\usepackage[normalem]{ulem}
\usepackage{subcaption}

\definecolor{bgray}{HTML}{F0F0F0}
\definecolor{bblue}{HTML}{E6F0FF}
\definecolor{bgreen}{HTML}{E8F5E9}
\definecolor{borange}{HTML}{FFF3E0}
\definecolor{HeadBG}{gray}{0.95} 
\definecolor{RowAlt}{gray}{0.98} 

\usepackage{amsmath, amssymb, bm}           
\usepackage{tabularx, booktabs, multirow, makecell, arydshln, array} 
\usepackage{enumitem}                        
\usepackage[x11names,dvipsnames,table]{xcolor} 
\usepackage{soul}                            
\usepackage[normalem]{ulem}                  
\usepackage[ruled,vlined]{algorithm2e}      
\usepackage{graphicx, subcaption}           
\usepackage{listings}                        
\usepackage{tcolorbox}                        
\tcbuselibrary{listings,skins,breakable}     
\usepackage[section]{placeins}               
\usepackage{listings}

\usepackage{xcolor}
\usepackage{listings}
\tcbuselibrary{listings,skins,breakable}
\usepackage[section]{placeins}

\usepackage{eso-pic}
\usepackage{graphicx}

\newcommand{\badgegray}[1]{\setlength{\fboxsep}{2.5pt}\colorbox{bgray}{\small\sffamily #1}}
\newcommand{\badgeblue}[1]{\setlength{\fboxsep}{2.5pt}\colorbox{bblue}{\small\sffamily #1}}
\newcommand{\badgegreen}[1]{\setlength{\fboxsep}{2.5pt}\colorbox{bgreen}{\small\sffamily #1}}
\newcommand{\badgeorange}[1]{\setlength{\fboxsep}{2.5pt}\colorbox{borange}{\small\sffamily #1}}

\newcommand{\LongCatHeader}{%
  \AddToShipoutPictureFG*{%
    \AtPageUpperLeft{%
      \raisebox{-1.9cm}[0pt][0pt]{%
        \hspace{2.5cm}%
        \includegraphics[height=2em]{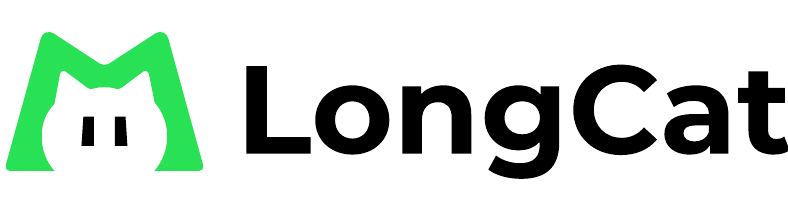}%
      }%
    }%
    \AtPageUpperLeft{%
      \raisebox{-2cm}[0pt][0pt]{%
        \hspace{2.5cm}%
        \rule{\dimexpr\paperwidth-5cm\relax}{1pt}%
      }%
    }%
  }%
}

\usepackage[final]{acl}

\usepackage{times}
\usepackage{latexsym}

\usepackage[T1]{fontenc}

\usepackage[utf8]{inputenc}

\usepackage{microtype}

\usepackage{inconsolata}

\usepackage{graphicx}
\usepackage{tikz}
\title{
Doctor-RAG: A Failure-Aware Repair Framework for Agentic Retrieval-Augmented Generation
}

\author{
\textbf{Shuguang Jiao}$^{1,4,}$\thanks{Work done during internship at Meituan.},
\textbf{Chengkai Huang}$^{2,3,}$\thanks{Corresponding authors.},
\textbf{Shuhan Qi}$^{1,}$\footnotemark[2],
\textbf{Xuan Wang}$^{1}$, \\[2pt]
\textbf{Yifan Li}$^{1}$,
\textbf{Quanchi Weng}$^{4}$,
\textbf{Lingchuan Liu}$^{4}$,
\textbf{Xunliang Cai}$^{4}$,
\textbf{Lina Yao}$^{3}$ \\[10pt]
$^{1}$Harbin Institute of Technology \quad
$^{2}$Macquarie University \quad
$^{3}$UNSW Sydney \quad
$^{4}$Meituan \\[2pt]
\texttt{24S151028@stu.hit.edu.cn} \quad
\texttt{shuhanqi@cs.hitsz.edu.cn}
}

\begin{document}

\maketitle

\LongCatHeader

\begin{abstract}
Agentic Retrieval-Augmented Generation interleaves retrieval and reasoning for multi-hop QA and complex knowledge tasks. As reasoning trajectories lengthen, failures become more frequent, while existing methods often either stop at diagnosis or rely on coarse replanning and rerun-style recovery, incurring high computational cost. We propose \textbf{Doctor-RAG (DR-RAG)}, a diagnose-and-repair framework that corrects failures via explicit error localization and prefix reuse. DR-RAG operates in two stages: (i) trajectory-level failure diagnosis, where a distilled diagnosis model jointly assesses evidence sufficiency, classifies the failure type, and localizes the earliest failure point; and (ii) tool-conditioned local repair that intervenes only at the diagnosed point while reusing conditionally valid prefixes and retrieved evidence. By separating error attribution from correction, DR-RAG avoids blind reruns in a post-hoc repair setting and enables targeted, efficient correction of known failed trajectories. Experiments on three multi-hop QA benchmarks across multiple agentic RAG baselines and backbone models show substantial improvements in answer accuracy. 
The code and data are available \href{https://github.com/Fdioa/dr_rag_open}{\uline{here}}.
\end{abstract}

\section{Introduction}

In recent years, Agentic Retrieval-Augmented Generation (Agentic RAG) has emerged as an important paradigm for multi-hop question answering and complex knowledge reasoning \citep{yao2023react,jin2025searchr1,song2025r1searcher}. By dynamically generating search queries, invoking retrievers, and performing multi-step reasoning during inference, Agentic RAG enables models to progressively plan and solve complex problems, improving evidence acquisition and multi-step problem-solving over static RAG systems \citep{shi2025deepresearchsurvey, singh2025artl, luo2025maragr1, zhang2025tearag}.

\begin{figure}[!t]
    \centering
\includegraphics[width=0.75\linewidth]{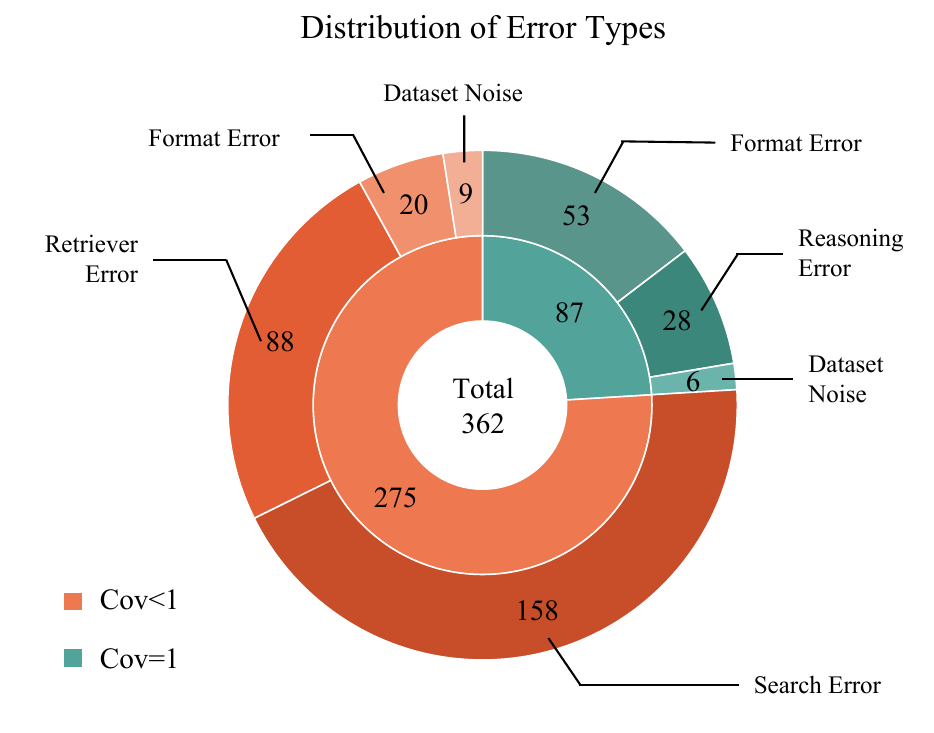}
    \vspace{-0.5em}
    \caption{Statistics of different error types in ReAct-based Agentic RAG on HotpotQA (500 samples).}
    \label{fig:error_distribution}
\end{figure}

However, as reasoning chains become longer and interaction rounds increase, failures in Agentic RAG become more frequent and difficult to avoid \citep{agrawal2024mindfulrag, leung2026classifying, ru2024ragchecker}. Since errors in multi-step reasoning may arise at arbitrary stages and accumulate over time, simply improving single-step reasoning or retrieval performance is insufficient to ensure stable system behavior \citep{dong2025ragcritic, yan2024corrective, DBLP:conf/iclr/AsaiWWSH24}. 
Therefore, the challenges faced by Agentic RAG extend beyond whether a correct answer can be generated, and increasingly concern:
\begin{quote}
\textbf{\textit{How can an Agentic RAG system localize the cause of a failure and repair it with minimal recomputation once the failure occurs?}}
\end{quote}
Existing studies have analyzed and modeled failures in RAG systems from different perspectives \citep{ru2024ragchecker, agrawal2024mindfulrag, leung2026classifying, cao2025outofstyle, guinet2024automated}. One line of work focuses on systematically categorizing and diagnosing errors in RAG outputs \citep{ru2024ragchecker, yehudai2025clear}, but does not further address automatic repair or performance improvement after failures occur. Another line of work uses diagnostic feedback to guide agentic repair \citep{dong2025ragcritic}, but typically follows a replanning--rerun strategy that discards the original trajectory and re-executes the retrieval--reasoning pipeline. These limitations suggest that failure analysis must be coupled with fine-grained recovery mechanisms that can selectively intervene in an existing trajectory rather than regenerate it from scratch.
Figure~\ref{fig:error_distribution} further illustrates this issue: failure types under the ReAct baseline on HotpotQA arise from diverse sources rather than a single dominant cause, making unconditional reruns or one-size-fits-all repair strategies inherently inefficient.

To address this problem, we propose \textbf{Doctor-RAG (DR-RAG)}, a unified diagnose-and-repair framework for Agentic RAG, designed to enable efficient failure correction through explicit error attribution and minimal-cost intervention. Unlike existing approaches that directly retry or replan the entire retrieval–reasoning pipeline, DR-RAG explicitly models failure correction as two consecutive stages: error diagnosis and conditional repair.

DR-RAG consists of two tightly coupled components: (1) \textbf{Taxonomy-Constrained Error Diagnosis and Localization.} When a reasoning failure occurs, DR-RAG treats the full agentic RAG trajectory as the object of diagnosis. A distilled diagnosis model jointly assesses evidence sufficiency, classifies the failure type under a sufficiency-aware taxonomy, and localizes the earliest failure point. This identifies the first reasoning or retrieval step that deviates from a valid solution path, providing structured signals for minimal intervention.  
(2) \textbf{Tool-Conditioned Local Repair.} Given the diagnosed error type and failure location, DR-RAG selects a repair operator that intervenes only at the failure point, while maximizing reuse of validated reasoning prefixes and retrieved evidence. Repair operations include lightweight answer rewriting for Format Error, evidence-preserving re-reasoning under full coverage, query rewriting and retrieval enhancement for Retriever Error, and planning-based repair for Search Error. By constraining repair to necessary components, DR-RAG reduces the cost of retrieval and reasoning while preserving coherence across repaired trajectories.

Our main contributions are as follows:
\begin{itemize}[leftmargin=1em, itemindent=0em, itemsep=0.3em]
    \item We are the first to formulate failure handling in Agentic RAG at the system level, modeling correction as diagnosis-guided minimal-cost intervention rather than full-pipeline retry.

    \item We propose a sufficiency-aware error taxonomy with a distilled diagnosis model that jointly performs evidence sufficiency assessment, failure classification, and earliest-step localization.

    \item We design tool-conditioned repair operators that adapt repair actions to diagnosed error types, enabling localized correction with prefix and evidence reuse.

    \item Experiments across multiple datasets, agentic RAG baselines, and backbone models show that DR-RAG achieves state-of-the-art repair rates while consistently improving downstream answer accuracy.

    
\end{itemize}

\section{Related Work}

\textbf{Retrieval-Augmented Generation.} LLMs are prone to hallucinations and outdated knowledge; RAG mitigates this by integrating external knowledge at inference time, improving factual accuracy without retraining \citep{guu2020retrieval, karpukhin2020dense,jiao2026prunerag,huang2025embedding}. The retriever–generator framework \citep{lewis2020retrieval} has been extended with advanced reasoning and retrieval, including Chain-of-Thought \citep{yao2023react}, Tree-of-Thought \citep{yao2023tree, jiao2026prunerag}, and adaptive retrieval \citep{jeong2024adaptive} for effective multi-step reasoning.

\textbf{Agentic RAG.} Agentic RAG (e.g., ReAct \citep{yao2023react}, Search-o1 \citep{li2025search}, Search-R1 \citep{jin2025searchr1}) unifies reasoning and tool use—retrieval \citep{dong2025toolstar}, memory \citep{li2025deepagent}, summarization \citep{dai2025evinoterag}—as sequential decision-making. Policy learning \citep{dong2025arpo, dong2025aepo} enables autonomous action selection; prior work \citep{wu2025hiprag, chen2025iterresearch, gao2025turns} optimizes policies via outcome-level rewards or preferences. Recent robustness studies include RAGChecker \citep{ru2024ragchecker} (evaluation-only) and RAG-Critic \citep{dong2025ragcritic} (trajectory replanning), but early error localization and prefix reuse remain underexplored, motivating our structured diagnosis and selective trajectory repair approach.

\textbf{Step-Level Online Verification and Repair.} 
Step-level online verification methods~\citep{lightman2023letsverify,xiong2024ipr,zhou2023lats} score intermediate actions or states during trajectory construction, enabling immediate correction but requiring access to the agent's internal execution loop. 
This makes them difficult to apply to black-box or closed-source agentic RAG systems, while frequent step-wise intervention may disrupt long-horizon reasoning and incur substantial per-step computation.
In contrast, DR-RAG avoids online step-wise intervention by diagnosing completed failed trajectories and applying localized repair with reuse of valid prefixes and evidence, making it compatible with black-box agentic RAG backbones while reducing repeated verification overhead.

\section{Agentic RAG Task Formulation}

We consider multi-step question answering under agentic retrieval-augmented generation (RAG), 
where a language model alternates between \textbf{Reason} ($r$), \textbf{Search} ($q$), 
and \textbf{Information} ($d$) actions.

Given a question $x \in \mathcal{X}$, the system produces an action trajectory:
\begin{equation}
y = \{ \rho^{(1)}, \rho^{(2)}, \dots, \rho^{(K)}, a_{\mathrm{pred}} \},
\end{equation}
where $\rho^{(k)} \in \{r, q, d\}$ and $a_{\mathrm{pred}}$ is the final answer (treated as a terminal action).

Actions typically interleave, and the trajectory length and structure may vary depending on intermediate reasoning outcomes, especially after localized repair, as illustrated in Figure~\ref{main_fig}(a).

\section{Methodology}

\subsection{DR-RAG Overview}

\begin{figure*}[t]
    \centering
    \includegraphics[width=\linewidth]{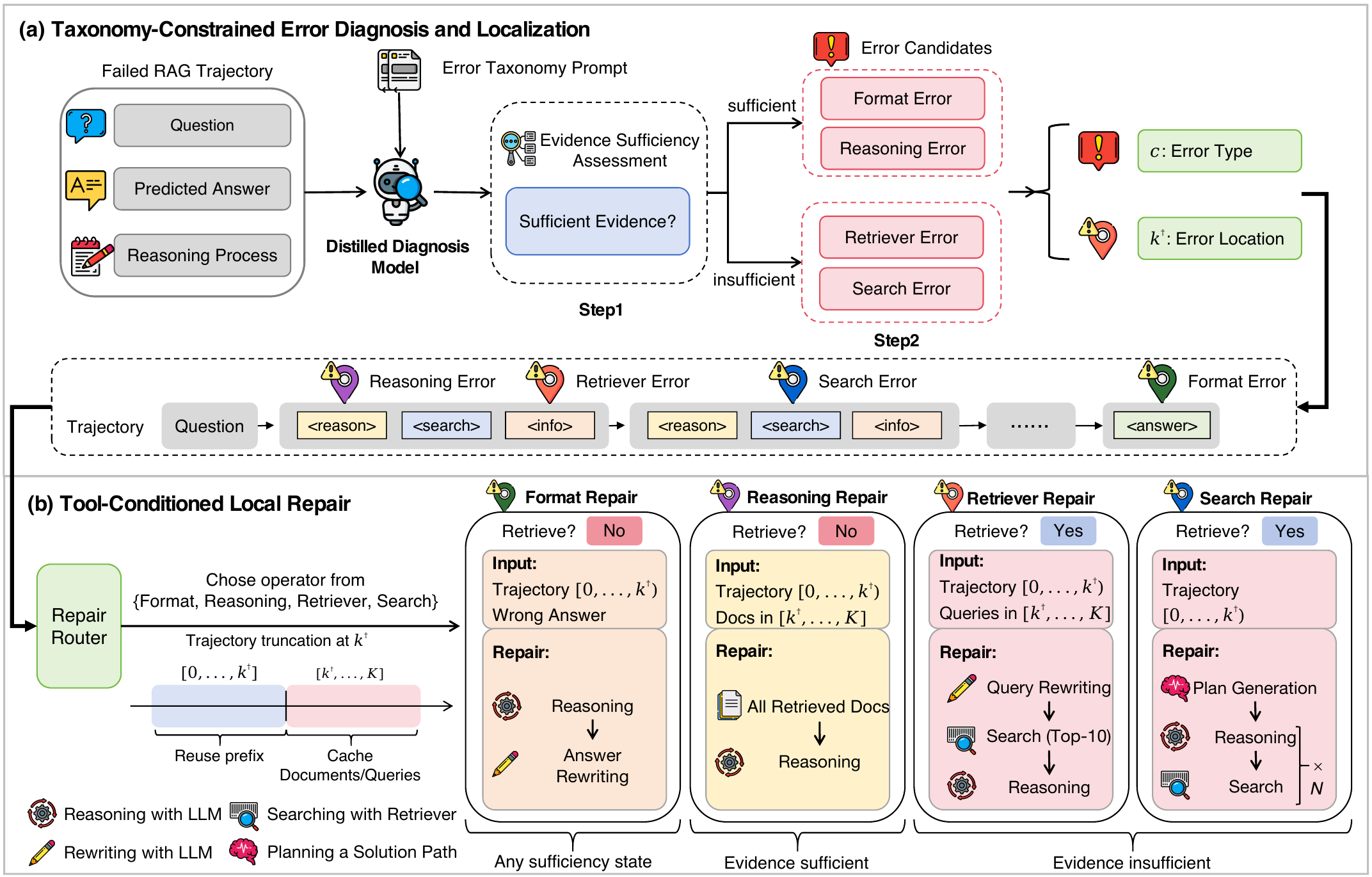}
    \vspace{-2em}
    \caption{Overview of DR-RAG. Given a failed agentic RAG trajectory, the diagnosis model jointly assesses evidence sufficiency and performs error classification and localization. Finally, tool-conditioned local repair intervenes at the localized failure point while reusing conditionally valid prefixes and retrieved evidence.}
    \label{main_fig}
\end{figure*}

Figure~\ref{main_fig} illustrates DR-RAG, a diagnose-and-repair framework for agentic RAG. 
Unlike prior approaches that rerun the full reasoning--retrieval pipeline, DR-RAG treats failed trajectories as reusable objects and corrects them via targeted, localized interventions. 

The framework operates in two stages: (1) error diagnosis and localization, and (2) tool-conditioned local repair.
During diagnosis, DR-RAG jointly assesses evidence sufficiency and identifies the error type and earliest failure point.
In repair, a specialized operator intervenes only at the failure point, reusing all previously validated prefix steps.

\begin{table}[t]
\centering
\resizebox{\columnwidth}{!}{
\begin{tabular}{l l l}
\toprule
\textbf{Evidence} & \textbf{Error Type} & \textbf{Localized Component} \\
\midrule
---
& Format Error
& \badgegray{answer} \\
\midrule
Sufficient
& Reasoning Error
& \badgegreen{reasoning} \\
\midrule
\multirow{2}{*}{Insufficient}
& Retriever Error
& \badgeblue{information} \\
& Search Error
& \badgeorange{search} \\
\bottomrule
\end{tabular}
}
\vspace{-1em}
\caption{Error taxonomy for agentic RAG trajectories with localized failure attribution. Evidence sufficiency is assessed implicitly by the diagnosis model.}
\label{tab:error_taxonomy}
\end{table}

\subsection{Diagnosis Module}
\label{sec:diagnosis_module}

The diagnosis module analyzes failed agentic RAG trajectories and outputs a structured diagnosis to condition downstream repair.
It takes as input a failed action-level trajectory $y$ produced by the underlying agentic RAG system, where $y$ contains reasoning steps, search queries, retrieved evidence blocks, observations, and a final predicted answer $a_{\mathrm{pred}}$.
The trajectory is considered failed when the prediction is incorrect, i.e., $\mathrm{EM}(a_{\mathrm{pred}}, a^\star)=0$.

The goal of diagnosis is to determine both \emph{why} the trajectory fails and \emph{where} the failure first occurs.
Formally, the module outputs a tuple $(c, k^\dagger)$, where $c$ denotes the predicted error type and $k^\dagger$ denotes the earliest localized failure point.
This structured diagnosis serves as the control signal for tool-conditioned local repair.

\subsubsection{Error Taxonomy}
\label{error_taxonomy}

Table~\ref{tab:error_taxonomy} summarizes the error taxonomy used by DR-RAG.
Each error type captures a repairable failure mode in the retrieval--reasoning process and restricts the downstream repair operator to a targeted intervention.
The taxonomy separates failures according to the earliest component that makes the trajectory unrecoverable without intervention.

\textbf{Format Error} captures cases where the final answer is semantically correct but fails exact match due to surface-form differences.
\textbf{Reasoning Error} captures cases where sufficient evidence has been retrieved, but the model derives an incorrect answer or intermediate conclusion.
\textbf{Retriever Error} captures cases where a valid search intent fails to retrieve the required evidence.
\textbf{Search Error} captures cases where the search itself is misdirected due to flawed reasoning, such as using incorrect entities, unsupported assumptions, wrong search directions, or overly vague queries.
Detailed definitions and decision criteria for each error type are provided in Appendix~\ref{app:diagnosis:criteria}.

\subsubsection{Decision Protocol and Localization}
\label{classification_localization}

The diagnosis follows a hierarchical protocol.
First, the model checks for \textbf{Format Error}, where the final answer is semantically correct but violates the exact-match format.
If the answer is incorrect, the model then assesses whether the retrieved evidence is sufficient to answer the question.
When the evidence is sufficient, the failure is classified as \textbf{Reasoning Error} and localized to the first reasoning step that deviates from a valid solution path.

When the evidence is insufficient, the model further distinguishes \textbf{Retriever Error} from \textbf{Search Error} through query-level validity judgment.
A search query is considered \emph{valid} if it uses correct entities from the question or prior evidence, targets relevant information, and is sufficiently specific.
If such a valid query fails to retrieve the necessary evidence, the error is attributed to a Retriever Error.
In contrast, a query is treated as \emph{poisoned} if it is caused by flawed reasoning, such as hallucinated entities, unsupported assumptions, wrong search directions, or excessive vagueness.
If missing evidence results from poisoned queries, the failure is classified as a Search Error.
If multiple errors occur along the same trajectory, the earliest occurring error determines the localized error type.

Localization is performed at the action level.
For Format Error, the failure point is the final answer step.
For Reasoning Error, it is the first reasoning step that produces an unsupported or incorrect intermediate conclusion despite sufficient evidence.
For Retriever Error, it is the evidence block following a valid query that fails to contain the necessary information.
For Search Error, it is the first poisoned search action---i.e., the earliest search query misdirected by flawed upstream reasoning.
Actions preceding $k^\dagger$ are treated as conditionally valid and can be reused during repair, while actions after $k^\dagger$ are considered unreliable.
Thus, the predicted error type $c$ determines which repair operator to invoke, and the localized point $k^\dagger$ determines where local repair starts. More detailed steps and the diagnostic prompt template are provided in Appendix~\ref{app:diagnosis}.

\subsubsection{Evidence Sufficiency in Diagnosis}

Evidence sufficiency is a key intermediate judgment in the above decision protocol, rather than an independent prediction module.
Instead of relying on a separate coverage predictor, DR-RAG estimates evidence sufficiency within the reasoning process of the diagnosis model.
After excluding format-invalid answers, the diagnostic prompt instructs the model to inspect the trajectory's \texttt{[Evidence]} blocks and determine whether they contain the necessary facts to support the correct answer.
Sufficient evidence indicates that the failure lies in how the model uses the evidence, leading to Reasoning Error.
Insufficient evidence indicates that the trajectory lacks necessary facts, and the diagnosis further determines whether this is caused by a retriever-side miss or by a reasoning-induced bad query.
This integrated design allows the diagnosis model to jointly reason about evidence sufficiency, error type, and failure location in a single inference pass, avoiding error propagation from an external coverage estimator.

\subsubsection{Distilled Diagnosis Model}

We train the diagnosis module via supervised fine-tuning (SFT). A high-capacity teacher model (Qwen3.5-Plus in thinking mode)~\cite{qwen35blog} annotates failed trajectories with structured diagnostic reasoning and compact labels covering all four error types.
The student model (Qwen3-8B~\citep{yang2025qwen3technicalreport}, LLaMA-3.1-8B-Instruct~\citep{grattafiori2024llama3herdmodels}) is trained to reproduce this reasoning in \texttt{<think>} tags and output a JSON object specifying the predicted error type $c$ and earliest failure point $k^\dagger$.
The teacher and student use the same prompt structure; the teacher's advantage lies in its stronger reasoning capability rather than additional information access.

\subsection{Repair Module}
Given the diagnosis output $(c, k^\dagger)$ from the diagnosis module, the repair stage performs intervention on the failed agentic RAG trajectory.
Rather than restarting the reasoning--retrieval process, DR-RAG applies a \emph{tool-conditioned repair operator} selected according to the diagnosed error type.
This enables partial reuse of conditionally valid trajectory prefixes and avoids redundant computation.

Formally, we define a set of repair operators:
$\mathcal{F} = \{ f_1, \ldots, f_L \},$
where each operator $f_\ell$ is specialized for repairing an error. Operator selection is diagnosis-driven:
$f = \mathcal{S}(c),$
where $\mathcal{S}: \mathcal{C} \rightarrow \mathcal{F}$ maps an error type $c$ to its repair operator.

\subsubsection{Repair under Sufficient Evidence}
When sufficient evidence is confirmed, repair corrects reasoning or answer generation without further retrieval.

\paragraph{Format-Invalid Answer Repair.}  
For format errors, the trajectory and retrieved documents are considered valid. Repair is limited to rewriting the final answer to meet task-specific constraints. The model is prompted with the original question, retrieved documents, and formatting instructions, and asked to regenerate the answer. No reasoning or retrieval is changed.

\paragraph{Reasoning Logic Error Repair.}  
For reasoning errors, the failure arises from incorrect reasoning over sufficient evidence. The trajectory is truncated at the localized error index $k^\dagger$, retaining the prefix:
\begin{equation}
y_{\mathrm{prefix}} = \{ \rho^{(1)}, \ldots, \rho^{(k^\dagger-1)} \},
\end{equation}
where $\rho^{(k)}$ denotes the $k$-th action.

Let $\mathcal{D}(y)$ denote document observation actions in $y$, and aggregate retrieved documents as:
\begin{equation}
D_{\mathrm{all}}(y) = \bigcup_{\rho \in \mathcal{D}(y)} \rho .
\end{equation}
The model re-performs reasoning from the retained prefix, grounded on $D_{\mathrm{all}}(y)$, discarding previous incorrect steps. This corrects reasoning errors without additional retrieval.

\begin{table*}[!t]
\centering
\footnotesize
\renewcommand{\arraystretch}{1}
\setlength{\tabcolsep}{4.2pt}
\resizebox{\textwidth}{!}{
\begin{tabular}{llc ccc ccc ccc}
\toprule
\multirow{2}{*}{\textbf{Baseline}} & \multirow{2}{*}{\textbf{Method}}  & \multirow{2}{*}{\textbf{\shortstack{Repair\\Rate}}($\uparrow$)} & \multicolumn{3}{c}{\textbf{HotpotQA}} & \multicolumn{3}{c}{\textbf{2Wiki}} & \multicolumn{3}{c}{\textbf{MuSiQue}} \\
\cmidrule(lr){4-6} \cmidrule(lr){7-9} \cmidrule(lr){10-12}
 & & &  \textbf{$\Delta\text{EM}$} & \textbf{$\Delta\text{F1}$} & \textbf{$\Delta\text{R-L}$} & \textbf{$\Delta\text{EM}$} & \textbf{$\Delta\text{F1}$} & \textbf{$\Delta\text{R-L}$} & \textbf{$\Delta\text{EM}$} & \textbf{$\Delta\text{F1}$} & \textbf{$\Delta\text{R-L}$} \\
 \midrule

\rowcolor{gray!20} \multicolumn{12}{c}{\textit{Qwen3-8B}} \\
\multirow{4}{*}{ReAct} & Rerun & 12.2 & 17.6 & 16.3 & 14.8 & 15.1 & 15.3 & 15.0 & 3.8 & 3.8 & 3.8 \\
 & Step-wise & 8.1 & 13.0 & 13.0 & 11.3 & 8.3 & 12.8 & 12.5 & 3.2 & 1.1 & 0.8 \\
 & RAG-Critic & 13.9 & 18.1 & 17.3 & 16.8 & 15.6 & 13.1 & 12.9 & 6.9 & 5.7 & 5.3 \\
 & DR-RAG & \textbf{21.8} & \textbf{31.6} & \textbf{27.3} & \textbf{26.9} & \textbf{26.0} & \textbf{24.4} & \textbf{24.5} & \textbf{7.7} & \textbf{6.2} & \textbf{6.2} \\
\midrule
\multirow{4}{*}{Search-o1} & Rerun & 9.8 & 14.7 & 11.8 & 12.2 & 7.3 & 5.2 & 5.1 & 7.3 & 3.5 & 3.5 \\
 & Step-wise & 7.3 & 12.7 & 9.1 & 6.4 & 5.6 & 1.0 & 0.2 & 3.5 & 0.5 & -0.2 \\
 & RAG-Critic & 9.9 & 13.2 & 12.0 & 12.6 & 5.6 & 2.0 & 2.1 & 10.8 & 9.8 & 9.8 \\
 & DR-RAG & \textbf{22.5} & \textbf{29.8} & \textbf{22.7} & \textbf{21.1} & \textbf{19.5} & \textbf{15.3} & \textbf{14.1} & \textbf{18.1} & \textbf{15.2} & \textbf{14.6} \\
\midrule
\multirow{4}{*}{Search-R1} & Rerun & 16.6 & 19.4 & 20.8 & 20.6 & 23.3 & 23.5 & 23.9 & 7.2 & 8.8 & 8.7 \\
 & Step-wise & 7.4 & 7.6 & 9.0 & 8.3 & 12.9 & 14.8 & 15.7 & 1.7 & 2.0 & 2.2 \\
 & RAG-Critic & 12.2 & 11.8 & 12.8 & 12.7 & 15.3 & 13.8 & 15.0 & 7.3 & 8.2 & 9.2 \\
 & DR-RAG & \textbf{19.8} & \textbf{25.4} & \textbf{26.1} & \textbf{25.1} & \textbf{25.8} & \textbf{24.4} & \textbf{25.3} & \textbf{8.3} & \textbf{9.0} & \textbf{9.7} \\

\rowcolor{gray!20} \multicolumn{12}{c}{\textit{LLaMA-3.1-8B-Instruct}} \\
\multirow{4}{*}{ReAct} & Rerun & 9.6 & 16.4 & 13.9 & 12.8 & 10.1 & 8.4 & 8.1 & 2.2 & 0.3 & -0.1 \\
 & Step-wise & 8.4 & 14.3 & 12.5 & 10.6 & 8.3 & 9.2 & 8.6 & 2.5 & 0.8 & 1.0 \\
 & RAG-Critic & 5.3 & 7.1 & 5.4 & 4.4 & 6.9 & 3.7 & 3.1 & 1.9 & -1.7 & -1.4 \\
 & DR-RAG & \textbf{15.5} & \textbf{27.8} & \textbf{23.4} & \textbf{21.3} & \textbf{13.3} & \textbf{9.9} & \textbf{9.5} & \textbf{5.4} & \textbf{1.5} & \textbf{1.5} \\
\midrule
\multirow{4}{*}{Search-o1} & Rerun & 7.6 & 12.7 & 7.8 & 6.7 & 5.6 & 0.8 & -0.0 & 4.5 & 2.3 & 2.1 \\
 & Step-wise & 6.8 & 9.3 & 4.0 & 1.6 & 6.7 & 1.7 & 0.7 & 4.5 & 1.4 & 0.7 \\
 & RAG-Critic & 5.7 & 9.8 & 7.2 & 7.2 & 5.6 & 2.4 & 2.3 & 1.7 & 0.1 & 0.2 \\
 & DR-RAG & \textbf{17.8} & \textbf{28.4} & \textbf{17.8} & \textbf{14.8} & \textbf{12.3} & \textbf{4.0} & \textbf{2.6} & \textbf{12.5} & \textbf{6.3} & \textbf{5.4} \\
\midrule
\multirow{4}{*}{Search-R1} & Rerun & 6.9 & 8.9 & 11.6 & 11.0 & 7.7 & 9.6 & 9.8 & 4.1 & 3.1 & 3.4 \\
 & Step-wise & 5.4 & 8.3 & 9.5 & 9.0 & 5.2 & 8.8 & 9.1 & 2.6 & 2.7 & 3.3 \\
 & RAG-Critic & 2.7 & 3.2 & 1.2 & 1.1 & 3.7 & 2.0 & 2.4 & 1.2 & -0.7 & -0.8 \\
 & DR-RAG & \textbf{12.4} & \textbf{17.2} & \textbf{14.9} & \textbf{14.1} & \textbf{13.5} & \textbf{10.0} & \textbf{10.6} & \textbf{6.5} & \textbf{4.5} & \textbf{3.8} \\

\bottomrule
\end{tabular}
}
\vspace{-1em}
\caption{Main repair results on three multi-hop QA benchmarks across two backbone LMs. 
All methods are evaluated on the same failed trajectories (EM=0). 
Repair Rate is averaged over datasets, while $\Delta$EM, $\Delta$F1, and $\Delta$R-L are reported per dataset. Best results within each baseline block are in \textbf{bold}.}
\label{tab:main_results}
\end{table*}

\subsubsection{Repair under Insufficient Evidence}
When retrieved evidence is insufficient, repair must address both retrieval failures and answer-generation issues.

\paragraph{Format-Invalid Answer Repair.}
Format errors may also occur under partial evidence coverage, since the model can sometimes infer the correct semantic answer from partial evidence or its parametric knowledge, even when not all supporting evidence is retrieved.
In this case, repair is limited to rewriting the final answer into the required form, without altering reasoning or retrieval steps.

\paragraph{Retriever Error Repair.}  
If the retriever fails to return relevant documents despite valid queries, the trajectory is truncated at $k^\dagger$ (the failed evidence block) and the prefix $y_{\mathrm{prefix}}$ is retained. The queries that produced insufficient evidence (at $k^\dagger$ and any subsequent failed retrievals), denoted $\mathcal{Q}_{\mathrm{failed}}$, are rewritten into multiple alternative formulations to improve recall while preserving intent. The retriever’s top-$k$ is increased to obtain more candidate documents, then the final answer is regenerated using the retained prefix and augmented evidence.

\paragraph{Search Error Repair.}  
If failure is caused by incorrect reasoning that misguides retrieval, reasoning, and retrieval beyond $k^\dagger$ are regenerated. After truncation, the model produces a high-level solution plan conditioned on the question and retained prefix, decomposing the task into reasoning and retrieval steps to generate a new trajectory suffix. Beyond the truncation point, both reasoning and retrieval may diverge from the original path.

\subsubsection{Summary of Repair Operators}  
All repair operators share a common structure: they preserve conditionally valid prefixes, intervene only at $k^\dagger$, and modify the minimal components required. This tool-conditioned design aligns with the diagnosis module, enabling efficient, localized repair.

\section{Experiments}

\subsection{Research Questions}

In this section, we address the following Research Questions (RQs):
\begin{itemize}[leftmargin=1em, itemindent=0em, itemsep=0.3em]
    \item \textbf{RQ1:} Does DR-RAG achieve higher repair rates and larger performance gains than other repair methods across different agentic RAG baselines?
    \item \textbf{RQ2:} Do error taxonomy and error localization each play an essential role in DR-RAG?
    \item \textbf{RQ3:} How efficient is DR-RAG during inference compared with other repair methods, when both diagnosis and repair costs are considered?
    \item \textbf{RQ4:} How accurately does the diagnosis model classify error types and localize failure points across different datasets and baselines?
    \item \textbf{RQ5:} How does DR-RAG's repair rate vary across different error types, and how does diagnosis correctness affect repair outcomes?
\end{itemize}

\subsection{Experiment Settings}

\textbf{Datasets and Metrics.} Experiments are conducted on three multi-hop QA benchmarks: HotpotQA~\citep{DBLP:conf/emnlp/Yang0ZBCSM18}, 2Wiki~\citep{DBLP:conf/coling/HoNSA20}, and MuSiQue~\citep{trivedi2022musique}, all providing gold supporting evidence for fine-grained analysis. 
We evaluate using standard QA metrics: Exact Match (EM), F1, and ROUGE-L~\citep{lin2004rouge}.  
To measure repair effectiveness, we report the \emph{Repair Rate}, defined as the proportion of initially incorrect instances (EM=0) corrected to exact match (EM=1).  
Absolute improvements relative to the unrepaired baseline are denoted as $\Delta$ (e.g., $\Delta\text{EM}$).
Detailed dataset statistics and metric definitions are provided in Appendix~\ref{app:datasets}.

\textbf{Baselines.} We evaluate DR-RAG on three agentic RAG baselines: \textbf{ReAct} \citep{yao2023react}, \textbf{Search-o1} \citep{li2025search}, and \textbf{Search-R1} \citep{jin2025searchr1}, all of which generate multi-step trajectories interleaving reasoning and retrieval. We compare DR-RAG with the following repair strategies applied to the same failed trajectories: 

\textbf{(1) Rerun} discards the entire failed trajectory and regenerates it from scratch, treating all prior steps as unreliable and unnecessarily recomputing prefixes that may still be valid.

\textbf{(2) Step-wise Retry} sequentially verifies reasoning steps and reuses the validated prefix for subsequent steps. While enabling partial reuse, it does not classify error types or explicitly diagnose the earliest failure point.

\textbf{(3) RAG-Critic} \citep{dong2025ragcritic} uses a diagnose--replan--rerun approach, identifying coarse error categories and replanning recovery actions. Diagnosis operates at the trajectory level without precise failure localization, which may lead to re-execution or modification of correct segments.

\textbf{Implementation Details.} Experiments follow prior agentic RAG work \citep{jiang-etal-2025-rag} using two backbone LMs: Qwen3-8B \citep{yang2025qwen3technicalreport}, LLaMA-3.1-8B-Instruct \citep{grattafiori2024llama3herdmodels}. All baselines and diagnosis-repair strategies share the same retrieval setup, 
using the full Wikipedia 2018 dump \citep{karpukhin2020dense}  indexed with FAISS \citep{johnson2021billion}, BGE-large-en-v1.5 \citep{xiao2023bge} as the retriever, and retrieving the top-5 documents for each query.
All inference costs incurred by diagnosis and repair are included in the efficiency evaluation. Experiments run on four NVIDIA RTX Pro 6000 GPUs using vLLM \citep{kwon2023efficient}. Full implementation details are provided in Appendix~\ref{app:setting}.

\subsection{Main Results (RQ1)}

Table~\ref{tab:main_results} reports repair performance across three benchmarks and two backbone LMs. 
DR-RAG consistently achieves the highest repair rate and metric gains across all settings, substantially outperforming competing repair strategies in most cases (e.g., 21.8\% vs.\ 13.9\% on ReAct with Qwen3-8B).

The advantage stems from taxonomy-guided localization: by identifying error type and the earliest failure step, DR-RAG applies targeted repair (format rewriting, re-reasoning, or re-retrieval) without regenerating valid prefixes. 
In contrast, Rerun often reproduces errors, Step-wise lacks error-type awareness, and RAG-Critic may over-repair due to imprecise localization.

DR-RAG also generalizes across backbones: on LLaMA-3.1-8B-Instruct, it achieves 12.4\%--17.8\% repair rates versus 2.7\%--9.6\% for competing methods, confirming that the approach is not tied to a specific backbone. 
Gains on MuSiQue are relatively smaller in several settings because the task is more challenging and leaves less recoverable room for post-hoc repair. 
Nevertheless, DR-RAG still consistently outperforms all baselines.

\subsection{Ablation Study (RQ2)}
\begin{table}[!t]
\footnotesize
\setlength{\tabcolsep}{2pt}
\resizebox{\columnwidth}{!}{
\begin{tabular}{lcccccc}
\toprule
\multirow{2}{*}{\textbf{Baseline}} & \multicolumn{2}{c}{\textbf{HotpotQA}} &
\multicolumn{2}{c}{\textbf{2Wiki}} & \multicolumn{2}{c}{\textbf{MuSiQue}} \\
\cmidrule(lr){2-3} \cmidrule(lr){4-5} \cmidrule(lr){6-7}
& \textbf{$\Delta\text{F1}$} & \textbf{$\Delta\text{EM}$} & \textbf{$\Delta\text{F1}$} &
\textbf{$\Delta\text{EM}$} & \textbf{$\Delta\text{F1}$} & \textbf{$\Delta\text{EM}$} \\
\midrule
\textbf{DR-RAG} & \textbf{25.4} & \textbf{28.9} & \textbf{21.4} & \textbf{23.8} & \textbf{10.1} &
\textbf{11.4} \\
\midrule
\textit{w/o Taxonomy} & 16.0 & 17.5 & 15.3 & 16.0 & 6.3 & 6.3 \\
\textit{w/o Localization} & 16.1 & 19.5 & 15.7 & 18.3 & 7.0 & 8.6 \\
\bottomrule
\end{tabular}%
}
\caption{Ablation study on error taxonomy and localization. Results are averaged over three baselines. Removing either component degrades both $\Delta$F1 and $\Delta$EM.}
\label{tab:ablation_study}
\end{table}

Table~\ref{tab:ablation_study} isolates the contributions of taxonomy and localization. Removing taxonomy (routing errors through a generic repair) drops $\Delta$EM by 7--11 points, as the model applies mismatched strategies (e.g., re-retrieval for format errors). Removing localization (repairing from the trajectory start rather than the failure point) causes a smaller but consistent drop of 5--9 points, due to unnecessary regeneration of correct prefixes. The two components are complementary: taxonomy selects \textit{what} to fix, while localization determines \textit{where} to intervene.

\begin{table}[!t]
\centering
\footnotesize
\resizebox{\columnwidth}{!}{
\begin{tabular}{llccc}
\toprule
\textbf{Baseline} & \textbf{Method} & \textbf{HotpotQA} & \textbf{2Wiki} & \textbf{MuSiQue} \\
\midrule
\multirow{4}{*}{ReAct} & Rerun  & 713  & 842 & 927 \\
& Step-wise & 1,031 & 915 & 990 \\
& RAG-Critic & 5,036 & 5,003  & 7,111 \\
& \textbf{DR-RAG}  & \textbf{308}  & \textbf{396} & \textbf{233} \\
\midrule
\multirow{4}{*}{Search-o1} & Rerun  & 633 & 716 & 817 \\
& Step-wise & 611  & 762 & 1,034 \\
& RAG-Critic & 4,735  & 5,595 & 8,790 \\
& \textbf{DR-RAG}  & \textbf{293}   & \textbf{403}  & \textbf{250} \\
\midrule
\multirow{4}{*}{Search-R1} & Rerun  & 652 & 474  & 773 \\
& Step-wise  & 530  & 702  & 734 \\
& RAG-Critic & 5,511  & 5,193  & 8,852 \\
& \textbf{DR-RAG}  & \textbf{289}  & \textbf{276} & \textbf{394} \\
\bottomrule
\end{tabular}%
}
\caption{Inference time (seconds) for repairing all failed trajectories per dataset. DR-RAG includes both diagnosis and repair time. Lowest values are in \textbf{bold}.}
\label{tab:efficiency}
\vspace{-0.8em}
\end{table}

\subsection{Inference Efficiency (RQ3)}

Table~\ref{tab:efficiency} compares end-to-end inference time across repair strategies. 
DR-RAG is 2--3$\times$ faster than Rerun and Step-wise, and over 10$\times$ faster than RAG-Critic, despite including the diagnosis overhead. 
The efficiency gain comes from localized repair: DR-RAG reuses conditionally valid prefixes and only performs generation for the diagnosed repair step, whereas Rerun regenerates entire trajectories and RAG-Critic incurs additional planning and execution overhead, as its generated repair plan may invoke multiple LLM-based repair operators.

\subsection{Error Diagnosis Reliability (RQ4)}

\begin{figure}[t]
    \centering
    \includegraphics[width=0.95\linewidth]{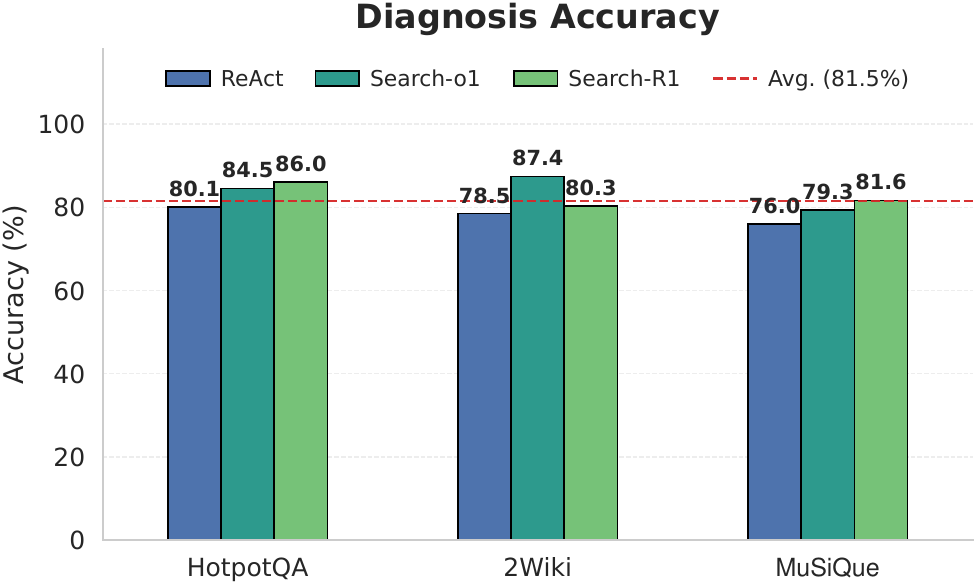}
    \caption{Accuracy of the automated diagnosis module evaluated against Qwen3.5-Plus-annotated diagnostic labels across different datasets and baselines.}
    \label{fig:repair_acc}
\end{figure}

\begin{figure}[t]
    \centering
    \begin{subfigure}[t]{0.48\linewidth}
        \centering
        \includegraphics[width=\linewidth]{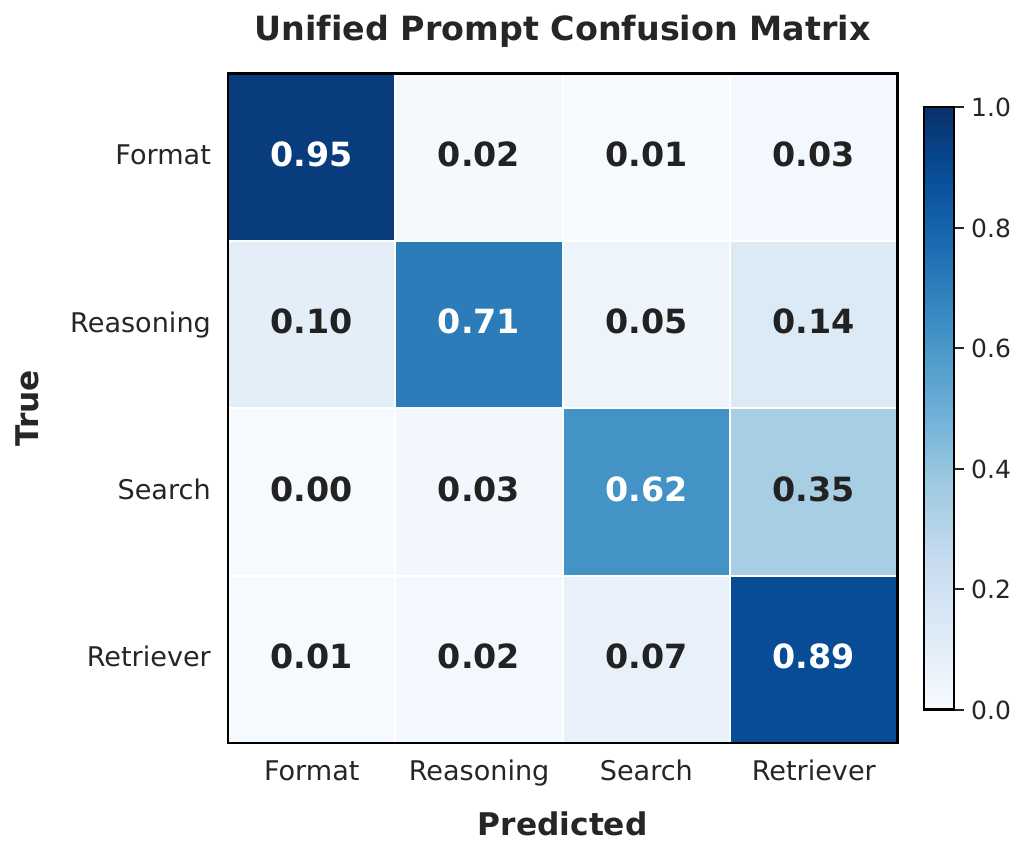}
        \caption{Confusion matrix for automated error diagnosis.}
        \label{fig:confusion}
    \end{subfigure}
    \hfill
    \begin{subfigure}[t]{0.48\linewidth}
        \centering
        \includegraphics[width=\linewidth]{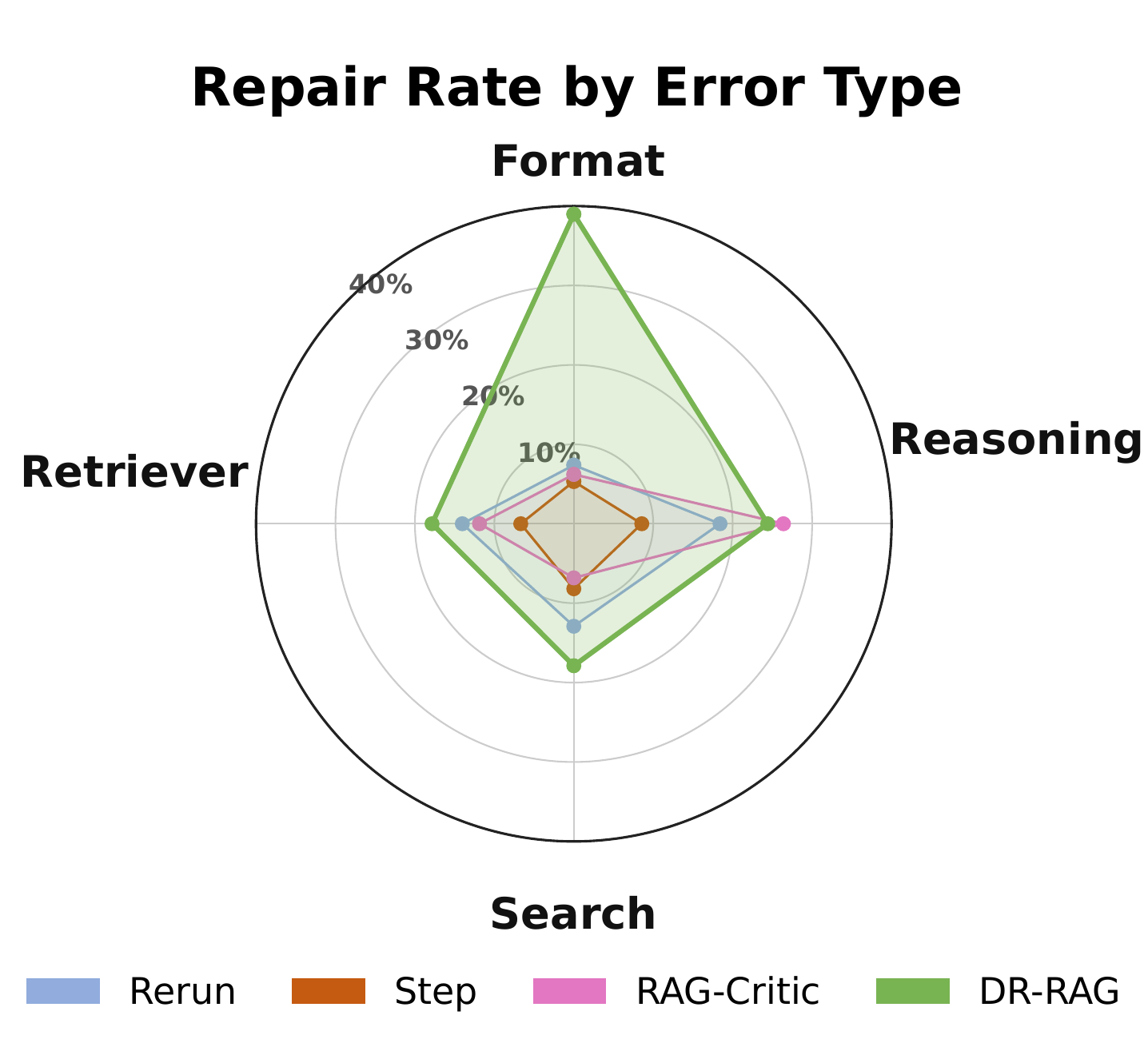}
        \caption{Average repair rates across error types.}
        \label{fig:radar}
    \end{subfigure}
    \caption{Automated error diagnosis and repair analysis.
(a) Confusion matrix aggregated across all datasets and baselines, showing high consistency between the diagnostic model and Qwen3.5-Plus labels.
    (b) Comparison of average repair rates across error types, aggregating results from three baselines under repair strategies.}
    \label{fig:diagnosis_and_repair}
\end{figure}

We evaluate DR-RAG's diagnosis module against Qwen3.5-Plus-annotated diagnostic labels across all 9 baseline$\times$dataset combinations (Figure~\ref{fig:repair_acc}). 
The module achieves 81.3\% overall accuracy, with per-configuration accuracy ranging from 76.0\% to 87.4\%, showing stable diagnosis reliability across datasets and baselines. 
This suggests that the distilled diagnosis model generalizes consistently across different agentic RAG settings rather than overfitting to a specific dataset or baseline.

Per-type analysis (Figure~\ref{fig:diagnosis_and_repair}a) shows that most error types are diagnosed reliably, especially Format (94.9\%) and Retriever (89.1\%) errors, which have clearer surface signals. 
Reasoning errors reach 70.7\% accuracy, while Search errors are lower at 62.0\% because they share the insufficient-evidence setting with Retriever errors and require finer causal attribution over query formation. 
Nevertheless, the impact is limited because Search errors are less frequent, and many misclassified cases are still routed to retrieval-enhanced repair, which can recover missing evidence through additional retrieval and re-reasoning. Additional robustness analysis is provided in Appendix~\ref{app:experiment:diagnosis_robustness}.

\subsection{Repair Performance by Error Type and Diagnostic Accuracy (RQ5)}
\begin{table}[!t]
\centering
\footnotesize
\setlength{\tabcolsep}{6pt}
\renewcommand{\arraystretch}{1.1}
\begin{tabular}{l c c c}
\toprule
\textbf{Error Type} & \textbf{Correct} & \textbf{Wrong} & \textbf{Overall} \\
\midrule
Format       & 41.3 & 11.1 & 39.5 \\
Reasoning    & 26.5 & 15.8 & 24.1 \\
Retriever    & 17.7 & 12.2 & 17.1 \\
Search       & 16.5 & 7.2 & 13.1 \\
\midrule
All          & 22.1 & 10.5 & 20.1 \\
\bottomrule
\end{tabular}
\caption{Repair rate (\%) by error type and diagnosis correctness. ``Correct''/``Wrong'' indicate whether the diagnosis matches the Qwen3.5-Plus label. Results are aggregated across baselines and datasets.}
\label{tab:repair_diag_avg}
\vspace{-1em}
\end{table}

Table~\ref{tab:repair_diag_avg} analyzes repair performance by error type and diagnostic correctness. 
Correct diagnoses consistently lead to higher repair rates than incorrect ones, confirming that accurate failure attribution is important for effective repair. 
At the same time, misdiagnosed cases still retain a non-trivial repair rate, indicating that the repair module is not brittle to imperfect diagnostic signals.

Across error types, Format and Reasoning errors are more repairable because they mainly require answer rewriting or re-reasoning over sufficient evidence, whereas Retriever and Search errors are harder since they require recovering missing evidence. 
Figure~\ref{fig:diagnosis_and_repair}b further visualizes this per-type repair trend across different repair strategies. 
Thus, RQ5 shows that DR-RAG is both diagnosis-sensitive and repair-stable: better diagnosis improves routing and recovery, while localized repair remains functional under diagnostic noise.

\section{Conclusion}

We introduce DR-RAG, a diagnose-and-repair framework for Agentic RAG that corrects failures via trajectory-level diagnosis and localized repair, avoiding full-pipeline reruns. 
Experiments demonstrate consistent improvements in accuracy and efficiency, supported by reliable automated diagnosis. 
DR-RAG reframes failure handling as a structured repair problem, providing a foundation for more robust and efficient multi-step reasoning systems.

\section*{Limitations}
While DR-RAG demonstrates strong trajectory-level diagnosis and localized repair for text-based Agentic RAG tasks, several natural extensions remain for future work.
First, the framework can be extended to multi-modal settings involving images, tables, or structured documents.
Second, DR-RAG can be combined with more diverse retrieval strategies and external knowledge sources.
Third, the repaired trajectories produced by DR-RAG may serve as improved supervision signals for trajectory distillation, enabling future systems to learn from corrected reasoning and retrieval behaviors.
Finally, applying DR-RAG to longer or more hierarchical reasoning trajectories may further broaden its applicability.
These directions do not affect the effectiveness of the proposed framework on the standard text-based tasks studied in this paper, but they provide useful opportunities for future exploration.

\bibliography{custom}

\appendix
\newpage
\onecolumn

\section*{Appendix Outline}
\label{appendix:outline}
{
\hypersetup{hidelinks}
}

\begin{figure*}[t]
    \centering
\includegraphics[width=1\linewidth]{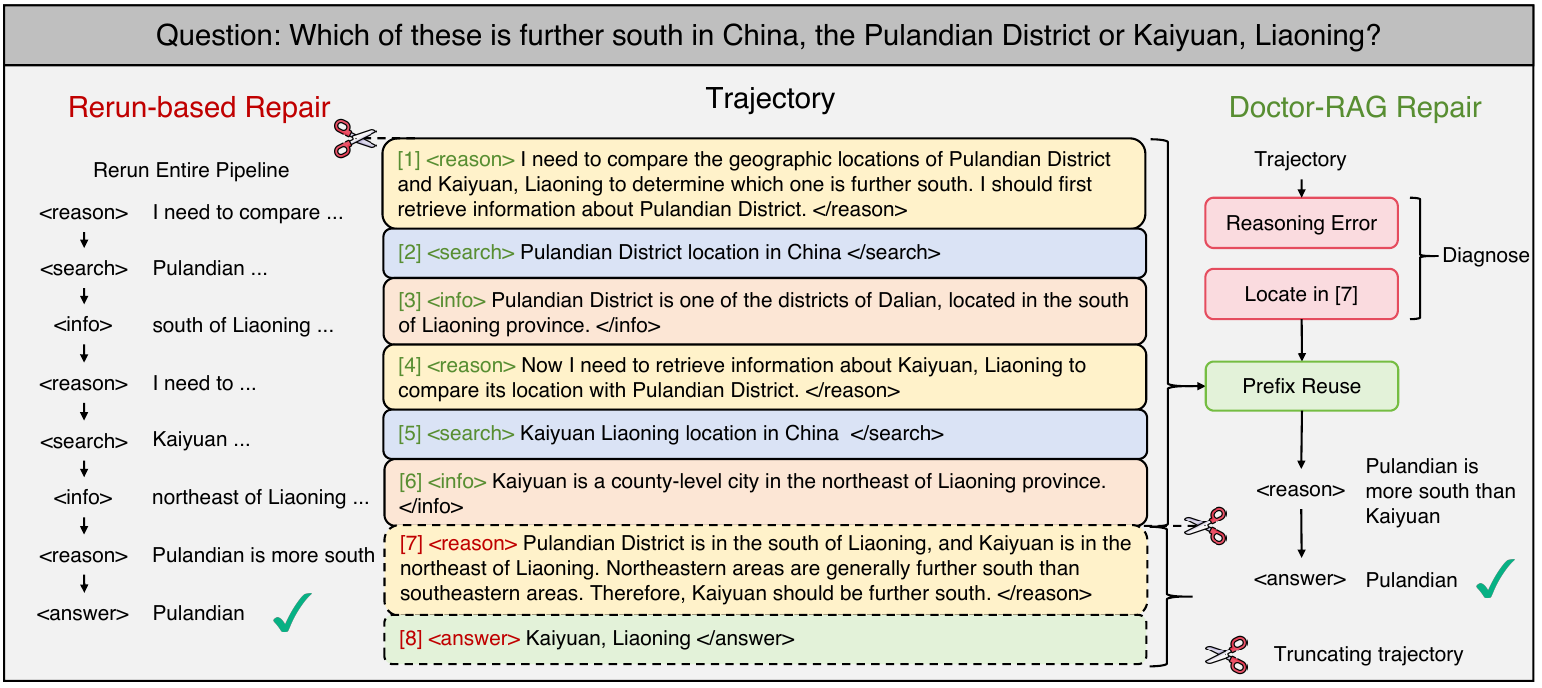}
    \caption{Case study of a reasoning logic error under full evidence coverage.
DR-RAG localizes the earliest faulty reasoning step and repairs the error by reusing retrieved evidence, avoiding unnecessary retrieval and full trajectory regeneration.}
    \label{case_fig}
\end{figure*}

\noindent
A.\,\hyperref[app:ai-assistants]{Information About Use of AI Assistants}\dotfill\pageref{app:ai-assistants}\\
B.\,\hyperref[app:impl]{Implementation Details}\dotfill\pageref{app:impl}\\
C.\,\hyperref[app:related_work]{Related Work}\dotfill\pageref{app:related_work}\\
\hspace*{1.5em}C.1\,\hyperref[app:related_work:rag]{Retrieval Augmented Generation}\dotfill\pageref{app:related_work:rag}\\
\hspace*{1.5em}C.2\,\hyperref[app:related_work:agentic_rag]{Agentic RAG}\dotfill\pageref{app:related_work:agentic_rag}\\
\hspace*{1.5em}C.3\,\hyperref[app:related_work:active_rag]{Comparison with Active and Step-Level Self-Corrective RAG}\dotfill\pageref{app:related_work:active_rag}\\
D.\,\hyperref[app:experiment]{Additional Experiments}\dotfill\pageref{app:experiment}\\
\hspace*{1.5em}D.1\,\hyperref[app:experiment:qwen4b]{Supplementary Main Results with Qwen3-4B}\dotfill\pageref{app:experiment:qwen4b}\\
\hspace*{3em}\hyperref[tab:main_results_suppl]{Supplementary Main Results Table}\dotfill\pageref{tab:main_results_suppl}\\
\hspace*{1.5em}D.2\,\hyperref[app:experiment:diagnosis_robustness]{Robustness Analysis of the Distilled Diagnosis Model}\dotfill\pageref{app:experiment:diagnosis_robustness}\\
\hspace*{3em}\hyperref[tab:diagnosis_accuracy]{Diagnosis Decomposition Table}\dotfill\pageref{tab:diagnosis_accuracy}\\
\hspace*{1.5em}D.3\,\hyperref[app:experiment:case]{Case Studies}\dotfill\pageref{app:experiment:case}\\
\hspace*{3em}\hyperref[case_fig]{Multiple Case Study Figures}\dotfill\pageref{case_fig}\\
E.\,\hyperref[app:diagnosis]{Diagnosis Algorithm Details}\dotfill\pageref{app:diagnosis}\\
\hspace*{1.5em}E.1\,\hyperref[app:diagnosis:pipeline]{Overall Diagnosis Pipeline}\dotfill\pageref{app:diagnosis:pipeline}\\
\hspace*{3em}\hyperref[alg:diagnosis]{Diagnosis Pipeline Algorithm}\dotfill\pageref{alg:diagnosis}\\
\hspace*{1.5em}E.2\,\hyperref[app:diagnosis:setup]{Problem Setup and Trajectory Representation}\dotfill\pageref{app:diagnosis:setup}\\
\hspace*{1.5em}E.3\,\hyperref[app:diagnosis:distillation]{Distilled Diagnosis Model}\dotfill\pageref{app:diagnosis:distillation}\\
\hspace*{3em}\hyperref[fig:diagnosis_output_schema]{Diagnosis Output Schema}\dotfill\pageref{fig:diagnosis_output_schema}\\
\hspace*{1.5em}E.4\,\hyperref[app:diagnosis:modes]{Diagnosis Logic}\dotfill\pageref{app:diagnosis:modes}\\
\hspace*{1.5em}E.5\,\hyperref[app:diagnosis:criteria]{Error Criteria and Query-Level Judgement}\dotfill\pageref{app:diagnosis:criteria}\\
\hspace*{3em}\hyperref[tab:error_criteria]{Error Type Definitions and Criteria}\dotfill\pageref{tab:error_criteria}\\
\hspace*{1.5em}E.6\,\hyperref[app:diagnosis:localization]{Failure Localization and Repair Interface}\dotfill\pageref{app:diagnosis:localization}\\
\hspace*{1.5em}E.7\,\hyperref[app:diagnosis:prompts]{Teacher and Student Prompts}\dotfill\pageref{app:diagnosis:prompts}\\
\hspace*{3em}\hyperref[app:prompt:unified]{Diagnosis Prompt}\dotfill\pageref{app:prompt:unified}\\
F.\,\hyperref[app:repair]{Repair Module Details}\dotfill\pageref{app:repair}\\
\hspace*{1.5em}F.1\,\hyperref[app:repair:format]{Format-Invalid Answer Repair}\dotfill\pageref{app:repair:format}\\
\hspace*{1.5em}F.2\,\hyperref[app:repair:reasoning]{Reasoning Logic Error Repair}\dotfill\pageref{app:repair:reasoning}\\
\hspace*{1.5em}F.3\,\hyperref[app:repair:retriever]{Retriever Error Repair}\dotfill\pageref{app:repair:retriever}\\
\hspace*{1.5em}F.4\,\hyperref[app:repair:search]{Search Error Repair}\dotfill\pageref{app:repair:search}\\
\hspace*{1.5em}F.5\,\hyperref[app:repair:algorithm]{Repair Algorithm Summary}\dotfill\pageref{app:repair:algorithm}\\
\hspace*{3em}\hyperref[alg:repair]{Tool-Conditioned Local Repair Algorithm}\dotfill\pageref{alg:repair}\\
\hspace*{1.5em}F.6\,\hyperref[app:repair:prompts]{Repair Prompt Templates}\dotfill\pageref{app:repair:prompts}\\
\hspace*{3em}\hyperref[app:prompt:format_repair]{Format-Invalid Answer Repair Prompt}\dotfill\pageref{app:prompt:format_repair}\\
\hspace*{3em}\hyperref[app:prompt:reasoning_repair]{Reasoning Logic Error Repair Prompt}\dotfill\pageref{app:prompt:reasoning_repair}\\
\hspace*{3em}\hyperref[app:prompt:query_rewrite]{Query Rewrite Prompt}\dotfill\pageref{app:prompt:query_rewrite}\\
\hspace*{3em}\hyperref[app:prompt:retriever_answer]{Answer Generation Prompt}\dotfill\pageref{app:prompt:retriever_answer}\\
\hspace*{3em}\hyperref[app:prompt:search_plan]{Search Plan Prompt}\dotfill\pageref{app:prompt:search_plan}\\
\hspace*{3em}\hyperref[app:prompt:search_execute]{Search Execution Prompt}\dotfill\pageref{app:prompt:search_execute}\\


\twocolumn
\newpage


\appendix

\section{Information About Use of AI Assistants}
\label{app:ai-assistants}

AI assistants were used during the research and writing process. Their use was limited to grammar checking, language polishing, improving manuscript readability, and assisting with part of the data annotation process. All annotations and manuscript revisions involving AI assistance were manually inspected and validated by the authors.

\section{Implementation Details}
\label{app:impl}
\subsection{Experimental Settings}
\label{app:setting}

Following prior work on agentic RAG systems \citep{jiang-etal-2025-rag}, we use two backbone language models in the main experiments: Qwen3-8B \citep{yang2025qwen3technicalreport} and LLaMA-3.1-8B-Instruct \citep{grattafiori2024llama3herdmodels}. 
We additionally include Qwen3-4B \citep{yang2025qwen3technicalreport} as a supplementary backbone setting in Appendix~\ref{app:experiment:qwen4b}. 
For each backbone setting, the corresponding diagnosis and repair models are instantiated with the same backbone. 
All baselines and diagnosis-repair strategies are evaluated under identical retrieval and inference settings to ensure fair comparison.
Following \citet{karpukhin2020dense}, we use the full Wikipedia 2018 dump as the retrieval corpus. Our main experiments follow the retrieval setup of \citet{jiang-etal-2025-rag}, employing FAISS \citep{johnson2021billion} for indexing, \textit{BGE-large-en-v1.5} \citep{xiao2023bge} as retriever, and retrieving top-5 documents. All methods share the same retrieval settings. All methods are evaluated in a post-hoc manner on failed trajectories produced by the original agentic RAG baselines, without modifying their underlying retrieval or reasoning components.
The evidence sufficiency indicator $\mathrm{Cov}(y)$ is estimated within the reasoning process of the diagnosis model, indicating whether the retrieved documents in a trajectory are sufficient to support the answer. 
This retrieval-sufficiency judgment is used to guide error classification and repair strategy selection.
We report inference time as the efficiency metric, including the time consumed by both diagnosis and repair.
Experiments are conducted on four NVIDIA RTX Pro 6000 GPUs, using vLLM for inference \citep{kwon2023efficient}.

\subsection{Datasets and Metrics}
\label{app:datasets}
We conduct experiments on three multi-hop question answering benchmarks: HotpotQA, 2Wiki, and MuSiQue.
HotpotQA~\citep{DBLP:conf/emnlp/Yang0ZBCSM18} is a large-scale benchmark for multi-hop reasoning over Wikipedia, 2WikiMultihopQA (2Wiki)~\citep{DBLP:conf/coling/HoNSA20} focuses on explicit multi-hop reasoning paths, and MuSiQue~\citep{trivedi2022musique} contains 2--4 hop questions constructed from multiple single-hop datasets.
All three datasets provide gold supporting evidence, enabling fine-grained analysis of evidence sufficiency and error diagnosis.

For evaluation, we report standard question answering metrics, including Exact Match (EM), F1, and ROUGE-L~\citep{lin2004rouge}. EM measures whether the predicted answer exactly matches the ground truth. F1 scores evaluate the token-level overlap, while ROUGE-L measures lexical similarity based on the longest common subsequence.

To evaluate error correction effectiveness, we report the \textit{Repair Rate}, defined as the proportion of instances that are initially incorrect (i.e., $\text{EM}=0$) but are successfully corrected to an exact match after repair (i.e., repaired $\text{EM}=1$).
For overall performance, we report the absolute improvement, denoted as $\Delta$, relative to the unrepaired baseline.
Specifically, $\Delta\text{EM}$ measures the net increase in the aggregate EM score attributable to the repair process.

\section{Related Work}
\label{app:related_work}
\subsection{Retrieval Augmented Generation}
\label{app:related_work:rag}

Large Language Models (LLMs) are prone to hallucinations and lack access to up-to-date knowledge. Retrieval-Augmented Generation (RAG) addresses these limitations by integrating external knowledge sources at inference time, improving factual accuracy without requiring model retraining \cite{guu2020retrieval, karpukhin2020dense}. Early work, exemplified by the original RAG framework \cite{lewis2020retrieval}, established the retriever–generator architecture that underpins retrieval-enhanced generation.
Building on this paradigm, subsequent studies progressively extend RAG with more advanced reasoning and retrieval mechanisms, including Chain-of-Thought \cite{yao2023react}, Tree-of-Thought reasoning \cite{yao2023tree, jiao2026prunerag}, and adaptive retrieval strategies \cite{jeong2024adaptive}. These advances enable effective multi-step reasoning in RAG systems.
\subsection{Agentic RAG}
\label{app:related_work:agentic_rag}
In recent years, Agentic RAG (e.g., ReAct \cite{yao2023react}, Search-o1 \cite{li2025search}, Search-R1 \cite{jin2025searchr1}) has formulated reasoning and tool invocation (retrieval \cite{dong2025toolstar}, memory \cite{li2025deepagent}, and summarization \cite{dai2025evinoterag}) as a unified sequential decision-making process. Through policy learning \cite{dong2025arpo, dong2025aepo}, these approaches enable models to autonomously select search and information processing actions during reasoning. Existing work \cite{wu2025hiprag, chen2025iterresearch, gao2025turns} typically optimizes reasoning and retrieval policies using outcome-level rewards or preference signals, thereby improving performance on multi-step search and complex reasoning tasks.

More recently, several studies have investigated robustness and failure diagnosis in RAG and agentic RAG systems. 
RAGChecker \cite{ru2024ragchecker} is a claim-level entailment-based diagnostic framework that evaluates RAG system performance but does not address how to repair or improve the system after errors are identified. 
RAG-Critic \cite{dong2025ragcritic} responds to detected error types by replanning a new reasoning and retrieval trajectory, without localizing the earliest error or reusing validated prefix trajectories.
As a result, failed trajectories are still treated as indivisible units, leading to redundant computation and limited error correction efficiency.

In contrast, fine-grained and cost-efficient repair of failed agentic RAG trajectories remains underexplored \cite{huang2026generative,huang2025listwise,lou2025speechagent}.
Prior work largely lacks explicit error localization within trajectories and does not consider reusing validated reasoning prefixes or retrieved evidence.
This gap motivates our work, which treats agentic RAG failures as structured objects that can be diagnosed, localized, and selectively repaired rather than blindly rerun.

\subsection{Comparison with Active and Step-Level Self-Corrective RAG}
\label{app:related_work:active_rag}

Self-corrective RAG methods improve reliability along two adjacent directions. 
The first is \emph{active, generation-time correction}, exemplified by Self-RAG~\citep{DBLP:conf/iclr/AsaiWWSH24} and CRAG~\citep{yan2024corrective}, where the model dynamically controls retrieval and evaluates evidence during generation using reflection tokens or lightweight evaluators. 
These methods mainly intervene in the ongoing generation process, deciding when to retrieve, critique, or continue generation. 
However, they do not target the post-hoc repair setting where a completed failed trajectory must be diagnosed, localized, and selectively corrected.

The second direction is \emph{step-level online verification}, where each action or intermediate state is scored using value functions, reward models, or external critics~\citep{lightman2023letsverify,xiong2024ipr,zhou2023lats}. 
Although such methods can provide immediate feedback, they require access to and modification of the agent's internal execution loop, making them difficult to apply to black-box agents or closed-source agentic RAG systems. 
Moreover, introducing verification at every step may disrupt the continuity of long-horizon reasoning, prematurely steer the model away from exploratory search paths, and impose substantial per-step computational overhead.

These two lines of work focus on improving reliability during trajectory construction, whereas DR-RAG studies how to recover from failures after a trajectory has already been generated. 
DR-RAG is designed as a post-hoc repair framework that treats the underlying agent as a black box. 
It only acts on failed trajectories, where a backbone-specific diagnosis model jointly assesses evidence sufficiency, classifies the failure type, and localizes the earliest erroneous step. 
Based on this structured diagnosis, DR-RAG applies localized repair while reusing valid reasoning prefixes and retrieved evidence. 
This design provides three key advantages: (i) it does not require modifying the underlying agent workflow, (ii) it reduces inference overhead by intervening only on failed trajectories, and (iii) it enables full-trajectory context for accurate failure localization and maximal prefix reuse.

Furthermore, DR-RAG's diagnosis module is trained using a \emph{distilled SFT} process: a strong teacher LLM annotates failed trajectories with structured reasoning and compact labels, while the student model learns to output the same diagnostic reasoning in a lightweight format. 
In contrast, online step-level verifiers typically require extensive per-step supervision, reward modeling, or iterative tool feedback during inference, which can be costly and tightly coupled to the agent workflow. 
By separating trajectory-level diagnosis from generation-time action selection, DR-RAG can be applied across different agentic RAG backbones without modifying or instrumenting the base agent, making it a flexible and efficient complement to existing self-corrective methods.





\begin{table*}[!t]
\centering
\footnotesize
\renewcommand{\arraystretch}{1}
\setlength{\tabcolsep}{4.2pt}
\resizebox{\textwidth}{!}{
\begin{tabular}{llc ccc ccc ccc}
\toprule
\multirow{2}{*}{\textbf{Baseline}} & \multirow{2}{*}{\textbf{Method}}  & \multirow{2}{*}{\textbf{\shortstack{Repair\\Rate}}($\uparrow$)} & \multicolumn{3}{c}{\textbf{HotpotQA}} & \multicolumn{3}{c}{\textbf{2Wiki}} & \multicolumn{3}{c}{\textbf{MuSiQue}} \\
\cmidrule(lr){4-6} \cmidrule(lr){7-9} \cmidrule(lr){10-12}
 & & &  \textbf{$\Delta\text{EM}$} & \textbf{$\Delta\text{F1}$} & \textbf{$\Delta\text{R-L}$} & \textbf{$\Delta\text{EM}$} & \textbf{$\Delta\text{F1}$} & \textbf{$\Delta\text{R-L}$} & \textbf{$\Delta\text{EM}$} & \textbf{$\Delta\text{F1}$} & \textbf{$\Delta\text{R-L}$} \\
 \midrule
\rowcolor{gray!20} \multicolumn{12}{c}{\textit{Qwen3-4B}} \\
\multirow{4}{*}{ReAct} & Rerun & 9.6 & 14.3 & 14.0 & 12.7 & 12.8 & 14.3 & 13.8 & 1.6 & 1.0 & 0.8 \\
 & Step-wise & 4.8 & 8.0 & 6.2 & 5.2 & 3.7 & 4.4 & 4.0 & 2.8 & -0.4 & -0.5 \\
 & RAG-Critic & 7.6 & 8.4 & 8.1 & 7.4 & 11.5 & 10.3 & 10.3 & 2.8 & 1.3 & 1.6 \\
 & \textbf{DR-RAG} & \textbf{18.1} & \textbf{28.2} & \textbf{24.2} & \textbf{22.5} & \textbf{18.8} & \textbf{16.8} & \textbf{17.1} & \textbf{7.4} & \textbf{6.2} & \textbf{6.8} \\
\midrule
\multirow{4}{*}{Search-o1} & Rerun & 8.2 & 12.7 & 10.7 & 9.4 & 7.8 & 5.5 & 5.5 & 4.2 & 1.8 & 1.5 \\
 & Step-wise & 3.9 & 6.4 & 2.1 & 0.6 & 3.4 & 0.8 & 0.4 & 2.1 & -1.0 & -1.9 \\
 & RAG-Critic & 7.3 & 13.7 & 10.9 & 12.0 & 3.9 & 2.6 & 2.5 & 4.2 & 3.9 & 4.1 \\
 & \textbf{DR-RAG} & \textbf{21.6} & \textbf{30.4} & \textbf{24.1} & \textbf{21.3} & \textbf{18.4} & \textbf{13.6} & \textbf{12.9} & \textbf{16.1} & \textbf{13.7} & \textbf{12.6} \\
\midrule
\multirow{4}{*}{Search-R1} & Rerun & 13.4 & 13.4 & 16.3 & 16.6 & 22.4 & 21.8 & 21.8 & 4.3 & 6.0 & 6.4 \\
 & Step-wise & 5.5 & 6.1 & 7.2 & 7.0 & 7.7 & 12.7 & 13.3 & 2.9 & 3.8 & 4.1 \\
 & RAG-Critic & 6.3 & 6.7 & 6.9 & 6.9 & 10.7 & 9.4 & 9.9 & 1.4 & 1.1 & 1.9 \\
 & \textbf{DR-RAG} & \textbf{17.6} & \textbf{20.1} & \textbf{22.8} & \textbf{22.6} & \textbf{26.1} & \textbf{26.0} & \textbf{26.4} & \textbf{6.7} & \textbf{8.4} & \textbf{8.3} \\
 
\bottomrule
\end{tabular}
}
\caption{Supplementary main results with Qwen3-4B as backbone. All methods operate on the same failed trajectories. The repair rate is averaged across three datasets. Best results per baseline are in \textbf{bold}.}
\label{tab:main_results_suppl}
\end{table*}

\section{Additional Experiments}
\label{app:experiment}

\subsection{Supplementary Main Results with Qwen3-4B}
\label{app:experiment:qwen4b}

Table~\ref{tab:main_results_suppl} reports supplementary results with Qwen3-4B, where both the diagnosis model and repair model are instantiated with Qwen3-4B. 
These results complement the main evaluation in Table~\ref{tab:main_results}, which uses Qwen3-8B and LLaMA-3.1-8B-Instruct.

DR-RAG achieves 18.1\%, 21.6\%, and 17.6\% repair rate on ReAct, Search-o1, and Search-R1, respectively, roughly doubling the best baseline in each case (Rerun: 9.6\%, 8.2\%, 13.4\%). The advantage is particularly large on Search-o1, where DR-RAG yields +30.4 $\Delta$EM on HotpotQA and +18.4 on 2Wiki, compared to +12.7 and +7.8 for Rerun. Step-wise retry performs worst among baselines (3.9\%--5.5\% RR), as the weaker backbone struggles to self-verify reasoning steps without explicit error-type guidance.

Compared to Qwen3-8B (Table~\ref{tab:main_results}), DR-RAG with Qwen3-4B shows only moderate degradation (e.g., 18.1\% vs.\ 21.8\% on ReAct), while baselines degrade more sharply (Rerun: 9.6\% vs.\ 12.2\%; RAG-Critic: 7.6\% vs.\ 13.9\%). This indicates that DR-RAG's structured diagnosis-then-repair paradigm is more robust to backbone capacity reduction than unstructured retry strategies, as the taxonomy and localization provide explicit guidance that compensates for weaker generation ability.



\subsection{Robustness Analysis of the Distilled Diagnosis Model}
\label{app:experiment:diagnosis_robustness}

A natural question for diagnosis-guided repair systems is how sensitive end-to-end performance is to diagnosis accuracy. We conduct a systematic robustness analysis across three dimensions: per-type diagnosis accuracy, stratified repair rates, and a controlled re-routing experiment.

\subsubsection{Diagnosis Accuracy Breakdown}

\begin{figure}[!t]
\centering
\includegraphics[width=1\columnwidth]{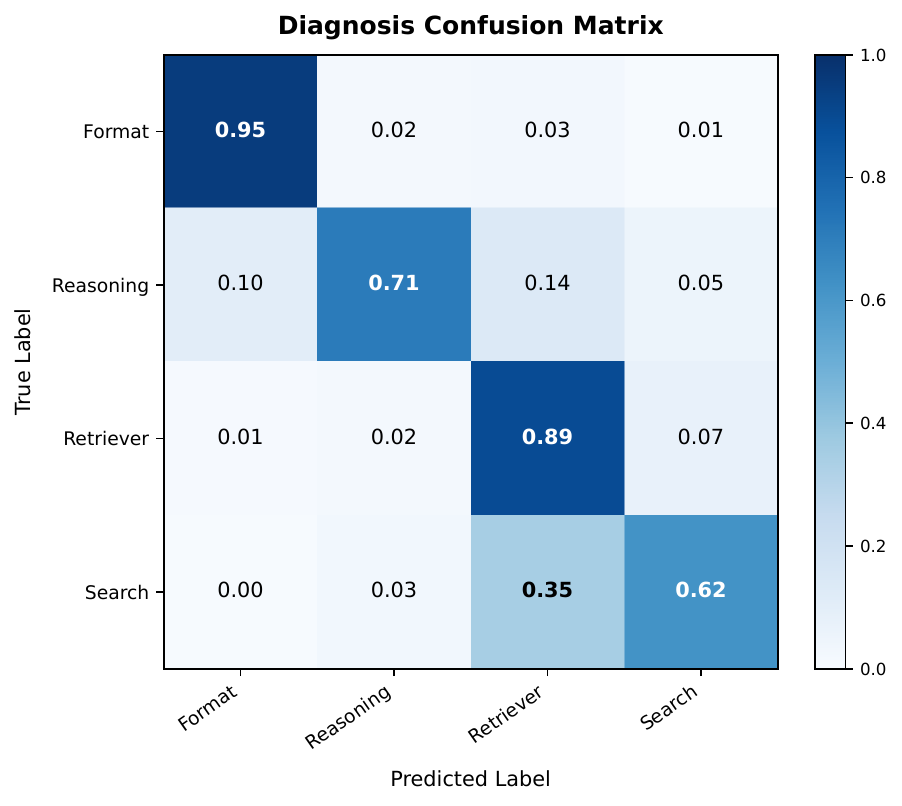}
\caption{Confusion matrix of the distilled diagnosis model (row-normalized). The primary confusion occurs between Retriever Error and Search Error.}
\label{fig:confusion_matrix}
\end{figure}

Table~\ref{tab:diagnosis_accuracy} reports the per-type classification accuracy of the distilled diagnosis model, evaluated against Qwen3.5-Plus-annotated \cite{qwen35blog} diagnostic labels across 3,131 valid samples (3 baselines $\times$ 3 datasets, excluding samples with invalid model outputs). The model achieves 81.3\% overall accuracy, with strong performance on Format Error (94.9\%) and Retriever Error (89.1\%). 
The primary source of confusion lies between Retriever Error and Search Error, whose boundary depends on distinguishing semantically appropriate queries that fail to retrieve sufficient evidence from queries flawed by upstream reasoning—a distinction that is inherently ambiguous.
Figure~\ref{fig:confusion_matrix} presents the full confusion matrix.

\begin{table}[!t]
\centering
\footnotesize
\setlength{\tabcolsep}{4pt}
\renewcommand{\arraystretch}{1.1}
\begin{tabular}{l r r r}
\toprule
\textbf{Error Type} & \textbf{Correct} & \textbf{Total} & \textbf{Accuracy} \\
\midrule
Format Error & 357 & 376 & 94.9\% \\
Reasoning Error & 297 & 420 & 70.7\% \\
Retriever Error & 1,455 & 1,633 & 89.1\% \\
Search Error & 435 & 702 & 62.0\% \\
\midrule
\textbf{Overall} & \textbf{2,544} & \textbf{3,131} & \textbf{81.3\%} \\
\bottomrule
\end{tabular}
\caption{Per-type diagnosis accuracy of the distilled diagnosis model evaluated against Qwen3.5-Plus-annotated diagnostic labels.}
\label{tab:diagnosis_accuracy}
\end{table}

\subsubsection{Stratified Repair Rates under Correct vs.\ Incorrect Diagnosis}

\begin{figure}[!t]
\centering
\includegraphics[width=1\columnwidth]{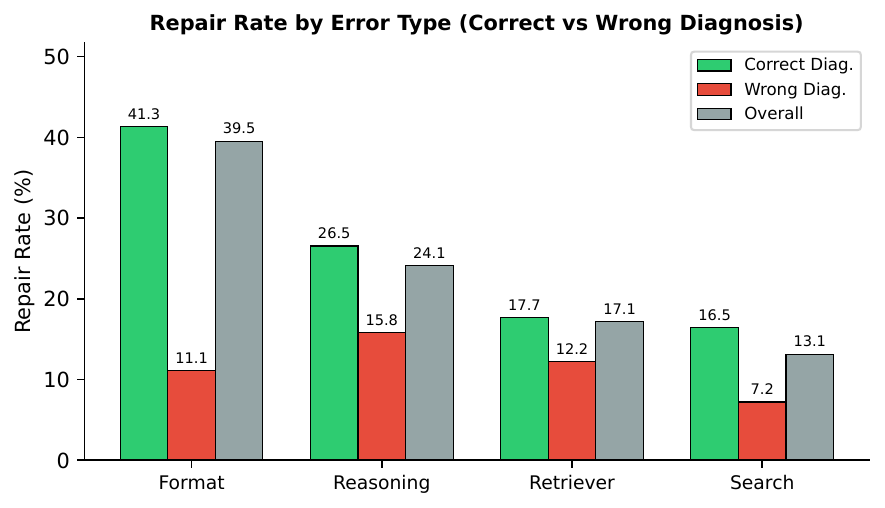}
\caption{Repair rates by error type, stratified by diagnostic correctness. Correct diagnosis consistently yields higher repair rates across all error categories.}
\label{fig:repair_by_type}
\end{figure}

To understand how diagnosis quality affects downstream repair, we stratify repair outcomes by whether the diagnosis was correct or incorrect (Table~\ref{tab:stratified_repair} and Figure~\ref{fig:repair_by_type}). Overall, correctly-diagnosed samples achieve a 22.1\% repair rate compared to 10.5\% for misdiagnosed samples, a gap of +11.7pp. Importantly, all four error types exhibit positive deltas, confirming that correct diagnosis consistently improves repair effectiveness.

The diagnosis advantage is most pronounced for Format Errors (+30.2pp), where correct identification enables targeted, lightweight answer reformatting. For Retriever Error and Search Error, the gap is smaller (+5.5pp and +9.3pp, respectively), suggesting that repair strategies for these adjacent categories have some functional overlap.

\begin{table}[!t]
\centering
\footnotesize
\setlength{\tabcolsep}{5pt}
\renewcommand{\arraystretch}{1.15}
\begin{tabular}{l c c c c}
\toprule
\textbf{Error Type} & \textbf{Diag. Acc.} & \textbf{Correct} & \textbf{Wrong} & \textbf{$\Delta$} \\
\midrule
Format & 94.9\% & 41.3\% & 11.1\% & +30.2 \\
Reasoning & 70.7\% & 26.5\% & 15.8\% & +10.7 \\
Retriever & 89.1\% & 17.7\% & 12.2\% & +5.5 \\
Search & 62.0\% & 16.5\% & 7.2\% & +9.3 \\
\midrule
\textbf{Overall} & \textbf{81.3\%} & \textbf{22.1\%} & \textbf{10.5\%} & \textbf{+11.7} \\
\bottomrule
\end{tabular}
\caption{Repair rates stratified by diagnosis correctness. ``Correct'' and ``Wrong'' denote the repair rates when the diagnosis model correctly or incorrectly classified the error type. $\Delta$ shows the gap in percentage points.}
\label{tab:stratified_repair}
\end{table}

\subsubsection{Controlled Re-routing Experiment}

We further isolate the causal effect of diagnosis errors through a controlled experiment. For the 587 misdiagnosed samples, we re-run the full repair pipeline with Qwen3.5-Plus-annotated error labels, holding all other variables constant---same repair model, retriever, and evidence context. Table~\ref{tab:rerouting} reports the results.

Correcting the routing for misdiagnosed samples yields a modest overall improvement (17.2\% $\rightarrow$ 19.8\%, +2.6pp). The improvement is most pronounced for Reasoning Errors (+5.7pp) and Format Errors (+5.3pp), where the correct repair strategy is substantially different from the misrouted one. For Retriever Error and Search Error, the gap is minimal (+2.2pp and +1.1pp), consistent with their shared query-rewriting repair paradigm. Notably, 62\% (364/587) of misroutes occur between Retriever and Search categories, explaining why the overall gain is modest despite correct routing being consistently beneficial.

\begin{table}[!t]
\centering
\footnotesize
\setlength{\tabcolsep}{5pt}
\renewcommand{\arraystretch}{1.15}
\begin{tabular}{l ccc}
\toprule
\textbf{Error Type} & \textbf{Wrong Route} & \textbf{Correct Route} & \textbf{n} \\
\midrule
Format & 10.5\% & 15.8\% & 19 \\
Reasoning & 33.3\% & 39.0\% & 123 \\
Retriever & 16.9\% & 19.1\% & 178 \\
Search & 10.5\% & 11.6\% & 267 \\
\midrule
\textbf{Overall} & \textbf{17.2\%} & \textbf{19.8\%} & \textbf{587} \\
\bottomrule
\end{tabular}
\caption{Controlled re-routing experiment on 587 misdiagnosed samples. “Wrong Route” shows repair rates under the original (incorrect) diagnosis; “Correct Route” shows rates after re-running repair with Qwen3.5-Plus-annotated labels.}
\label{tab:rerouting}
\end{table}

\subsubsection{Discussion}

Our analysis reveals several important insights into the behavior and robustness of the DR-RAG framework.

First, correct diagnosis provides a substantial and consistent benefit: across all error types, correctly-diagnosed samples achieve higher repair rates (+11.7pp overall, Table~\ref{tab:stratified_repair}). This validates the core premise of diagnosis-guided repair---accurate error classification enables targeted repair strategies that outperform generic approaches.

Second, the controlled re-routing experiment (Table~\ref{tab:rerouting}) confirms that correct routing improves outcomes, but the gain is modest (+2.6pp overall). This demonstrates \emph{graceful degradation}---the system maintains meaningful repair even under imperfect diagnosis. The modest gap can be explained by three factors:

\begin{itemize}[leftmargin=1em, itemindent=0em, itemsep=0.3em]
    \item \textbf{Redundancy in repair strategies:} Retriever Error and Search Error share a common evidence-augmented repair paradigm (query rewriting + re-retrieval). 62\% of misroutes between these categories still execute functionally similar repair operations.
    \item \textbf{Cross-effectiveness of repair strategies:} Some repair strategies are partially effective for multiple error types. For example, the retrieval-augmented repair can sometimes fix reasoning issues by providing better evidence.
    \item \textbf{Targeted benefit for distinct categories:} The routing correction benefit is concentrated in Reasoning (+5.7pp) and Format (+5.3pp) errors, where the repair strategies are fundamentally different from the misrouted ones.
\end{itemize}

These factors collectively demonstrate that DR-RAG achieves robust performance even when the finer-grained distinction between errors is incorrect. At the same time, precise diagnosis remains valuable---particularly for Format Errors, where correct classification increases the repair rate by +30.2pp (Table~\ref{tab:stratified_repair}), and for Reasoning Errors, where correct routing yields +5.7pp improvement (Table~\ref{tab:rerouting}). This highlights the architectural design choice of grouping functionally related error types under shared repair paradigms, ensuring resilience to diagnostic errors while maximizing the benefit of accurate classification in critical scenarios.

\subsection{Case Study}
\label{app:experiment:case}
We present a representative example from HotpotQA to illustrate how DR-RAG repairs reasoning errors under full evidence coverage.
The question asks: \emph{“Which of these is further south in China, the Pulandian District or Kaiyuan, Liaoning?”}
The agent retrieves the correct documents for both locations, indicating full evidence coverage.
However, it incorrectly answers \emph{“Kaiyuan, Liaoning”}, despite the retrieved evidence stating that Pulandian District is located in the south of Liaoning province, while Kaiyuan is in the northeast.

According to our taxonomy, this failure is classified as a reasoning logic error under full coverage, where the error originates from incorrect reasoning rather than missing evidence.
Restart-based baselines respond to this failure by regenerating the entire retrieval–reasoning trajectory, even though the retrieved documents contain sufficient information to answer the question.

In contrast, DR-RAG diagnoses the earliest reasoning failure and performs localized repair by reusing the retrieved evidence and recomputing only the affected reasoning steps.
This targeted intervention corrects the error without additional retrieval and avoids unnecessary recomputation, demonstrating the effectiveness of diagnosis-guided localized repair.


\section{Diagnosis Algorithm Details}
\label{app:diagnosis}

This appendix provides the detailed implementation of the DR-RAG diagnosis module.
DR-RAG uses a backbone-specific diagnosis model that jointly performs evidence sufficiency assessment, error classification, and failure localization.
Rather than relying on a separate coverage predictor to gate the candidate error space, DR-RAG integrates evidence sufficiency reasoning directly into the diagnostic prompt, allowing the model to implicitly determine whether retrieved evidence is sufficient as part of its classification process.

For each backbone setting, the diagnosis model is fine-tuned via SFT to output a structured diagnosis including both the error type and the localized failure point.
The prompt template defines the diagnostic protocol---including the evidence sufficiency check as a branching point---and the output format, while the final classification and localization are produced by the fine-tuned student model.

We first present the complete diagnosis pipeline in Algorithm~\ref{alg:diagnosis}, which connects teacher annotation, student distillation, inference-time diagnosis, and repair routing.
We then describe the trajectory representation, the diagnosis logic with its four error types, admissible behavior for each component, error-type criteria, query-level judgement, and the failure localization interface.
The key innovation is that the model itself determines evidence sufficiency: if retrieved evidence is sufficient but reasoning fails, the error is \textit{Reasoning Error}; if evidence is insufficient due to retriever failure or flawed query generation, the error is \textit{Retriever Error} or \textit{Search Error}, respectively.

\subsection{Overall Diagnosis Pipeline}
\label{app:diagnosis:pipeline}

Algorithm~\ref{alg:diagnosis} summarizes the complete diagnosis pipeline of DR-RAG.
The pipeline consists of two stages.
The first stage constructs SFT supervision using a teacher model that receives the ground-truth answer but not golden evidence.
The second stage deploys the distilled student model with the same prompt structure (including the ground-truth answer) to predict the error type and localized failure point.
A key design choice is that the teacher and student prompts are fully aligned: both receive the question, trajectory, and ground-truth answer, and follow the same diagnostic logic.
The teacher's advantage lies only in its stronger reasoning capability (Qwen3.5-Plus in thinking mode), not in additional oracle information.

\begin{algorithm*}[t]
\small
\caption{Diagnosis Pipeline of DR-RAG}
\label{alg:diagnosis}
\KwIn{
Failed trajectory set $\mathcal{D}$ for training;
a failed test trajectory $(q, y)$ for inference.
}
\KwOut{
Diagnosis result $(c, k^\dagger)$, where $c$ is the error type and $k^\dagger$ is the localized failure point.
}

\BlankLine
\textbf{Stage I: Teacher Annotation and Distillation}\;
Initialize SFT dataset $\mathcal{S}\leftarrow\emptyset$\;

\ForEach{failed training instance $d=(q, y, a^*)\in\mathcal{D}$}{
    Parse trajectory $y$ into indexed reasoning steps, search actions, evidence blocks, and final answer\;

    Construct the diagnosis prompt with $(q, y, a^*)$\;

    Run the teacher model to obtain a diagnostic rationale $r$ and structured label $j$\;

    Construct one SFT example: input = diagnosis prompt with $(q, y, a^*)$; output = $\langle$think$\rangle r \langle$/think$\rangle j$\;

    Add the example to $\mathcal{S}$\;
}

Fine-tune the student diagnosis model on $\mathcal{S}$\;

\BlankLine
\textbf{Stage II: Student Diagnosis at Inference Time}\;
Parse the failed test trajectory $y$ into indexed components\;

Construct the diagnosis prompt with $(q, y, a^*)$\;

The student performs a diagnosis following the prompt's hierarchical logic:\

\textbf{Phase 1:} Check if final answer is semantically equivalent to GT $\Rightarrow$ \textit{Format Error}\;

\textbf{Phase 2:} Assess evidence sufficiency from the trajectory's [Evidence] blocks:\;
\eIf{evidence is sufficient to derive the GT answer}{
    \textbf{Phase 3:} Locate the earliest divergent reasoning step $\Rightarrow$ \textit{Reasoning Error}\;
}{
    \textbf{Phase 4:} Perform entity provenance check on each search query\;

    \eIf{$\exists$ valid query with failed retrieval}{
        Set $c$ to \textit{Retriever Error} and $k^\dagger$ to the failed evidence block\;
    }{
        Set $c$ to \textit{Search Error} and $k^\dagger$ to the first poisoned search action\;
    }
}

\If{$c$ is \textit{Format Error}}{
    Set $k^\dagger\leftarrow\texttt{null}$\;
}

Return $(c, k^\dagger)$ as the control signal for local repair\;
\end{algorithm*}

\subsection{Problem Setup and Trajectory Representation}
\label{app:diagnosis:setup}

Given a question $q$, an agentic RAG system produces a trajectory $y$ that interleaves reasoning steps, search actions, retrieved evidence, and the final answer.
DR-RAG performs diagnosis on failed trajectories where the original answer does not match the ground-truth answer under exact match evaluation, i.e., $\mathrm{EM}=0$.

To support fine-grained fault localization, the raw trajectory is parsed into indexed components:
\begin{itemize}[nosep,leftmargin=1.5em]
    \item \texttt{[Step $k$] (Reasoning)}: an intermediate reasoning or planning step;
    \item \texttt{[Action $j$] (Search)}: a retrieval query issued by the agent;
    \item \texttt{[Evidence $j$] (Information)}: the retrieved content returned by the retriever (paired with Action~$j$);
    \item \texttt{[Step $k$] (Answer)}: the final answer generation step.
\end{itemize}

Reasoning and answer steps share the index $k$, while search actions use a separate action index $j$.
This representation allows the diagnosis model to return a culprit step label such as \texttt{culprit\_step\_label = "[Step 3] (Reasoning)"} or \texttt{culprit\_step\_label = "[Action 2] (Search)"}.
The indexed trajectory also provides the repair module with a concrete intervention point, enabling reuse of the valid prefix before the localized failure.

\subsection{Distilled Diagnosis Model}
\label{app:diagnosis:distillation}

The diagnosis module is implemented as a distilled diagnosis model.
Specifically, for each backbone setting, we fine-tune the corresponding student diagnosis model via supervised fine-tuning (SFT) to perform error classification and failure localization.
The training process follows a teacher--student distillation pipeline.

First, a stronger teacher model, Qwen3.5-Plus in thinking mode, annotates failed trajectories using the same diagnostic prompt as the student.
During annotation, the teacher receives the question, parsed trajectory, and ground-truth answer---the same input as the student.
The teacher does \emph{not} receive golden evidence or golden context.
Its advantage over the student lies solely in its stronger reasoning capability, which produces higher-quality diagnostic rationales.
This prompt-aligned design ensures that the distilled reasoning traces are directly compatible with the student's inference-time input, avoiding information leakage.

Second, the teacher output is converted into SFT data.
Each SFT example uses the diagnostic prompt (with question, trajectory, and ground-truth answer) as input.
The output contains the teacher's diagnostic rationale (extracted from thinking mode) followed by a compact JSON label.
Figure~\ref{fig:diagnosis_output_schema} illustrates the structured output schema.

\begin{figure}[t]
\centering
\begin{tcolorbox}[
  enhanced,
  colback=gray!3,
  colframe=gray!50!black,
  arc=4pt,
  boxrule=0.6pt,
  width=\columnwidth,
  title=\textbf{Diagnosis Output Schema},
  coltitle=black,
  colbacktitle=gray!12,
  boxed title style={colframe=gray!50!black,colback=gray!12,arc=4pt},
  attach boxed title to top left={xshift=6pt,yshift*=-\dimexpr\tcboxedtitleheight/2\relax},
  fontupper=\footnotesize,
  top=4pt,bottom=4pt,left=6pt,right=6pt
]

\textbf{Output} (all error types):
\begin{tcolorbox}[
  colback=white,
  colframe=gray!50,
  arc=2pt,
  boxrule=0.4pt,
  left=4pt,right=4pt,top=3pt,bottom=3pt
]
\texttt{\{"assigned\_error\_type": "...",}\\
\texttt{\ \ "culprit\_step\_label": "..."\}}
\end{tcolorbox}

\vspace{0.3em}
{\footnotesize
\texttt{assigned\_error\_type} $\in$ \{Format-Invalid Answer Error, Reasoning Logic Error, Retriever Failure, Reasoning-Induced Missing Retrieval\}.\\
\texttt{culprit\_step\_label}: the label of the first erroneous step (e.g., ``[Step 2] (Reasoning)'', ``[Action 1] (Search)'', or ``[Evidence 2] (Information)''). Set to \texttt{null} for Format Error.
}

\end{tcolorbox}
\caption{Structured output schema of the diagnosis model. A single format covers all four error types.}
\label{fig:diagnosis_output_schema}
\end{figure}

Third, the student model is trained to reproduce the teacher's diagnostic behavior.
At inference time, the student receives the question $q$, trajectory $y$, and ground-truth answer $a^*$, and outputs the predicted error type and failure location.
This design separates training-time distillation (where the teacher's stronger reasoning produces reliable labels) from inference-time diagnosis (where the student applies the same prompt independently).

We train the student model with full-parameter SFT using LLaMA-Factory \citep{zheng2024llamafactory}.
The learning rate is $5\times10^{-6}$ with a cosine scheduler and 10\% warmup.
We train for 3 epochs with batch size 1, gradient accumulation 16, cutoff length 16384, DeepSpeed ZeRO-2, bf16 precision, and \texttt{enable\_thinking=true}.

\subsection{Diagnosis Logic}
\label{app:diagnosis:modes}

The diagnosis model covers all four error types within a single prompt.
Rather than relying on an external coverage signal to gate the candidate error space, the model itself assesses evidence sufficiency as an intermediate reasoning step.
This evidence sufficiency check serves as the key branching point in the diagnostic logic.

The diagnostic process follows a strict priority order:

\paragraph{Phase 1: Format Error.}
The model first checks whether the final answer is semantically equivalent to the ground-truth answer but fails exact match due to surface-form mismatch (e.g., ``27'' vs.\ ``27 years old'', ``Steve Jobs'' vs.\ ``Steven Paul Jobs'').
If so, the error is classified as \textit{Format Error}.

\paragraph{Phase 2: Evidence Sufficiency Check.}
If the answer is factually wrong, the model assesses whether the trajectory's \texttt{[Evidence]} blocks contain sufficient information to derive the ground-truth answer.
This is the key branching point:
\begin{itemize}[nosep,leftmargin=1.5em]
    \item \textbf{Evidence sufficient}: the model had the right information but failed to use it correctly $\Rightarrow$ proceed to Phase~3.
    \item \textbf{Evidence insufficient}: the model was missing critical information $\Rightarrow$ proceed to Phase~4.
\end{itemize}

\paragraph{Phase 3: Reasoning Error.}
When evidence is sufficient, the failure must lie in the reasoning process---specifically, the model reasons incorrectly when \emph{consuming} already-sufficient evidence (e.g., misinterpreting a retrieved fact or ignoring a key passage).
The model locates the earliest step where reasoning diverges: hallucination, contradiction, wrong entity/date, misinterpretation, or ignored evidence.

\paragraph{Phase 4: Retrieval Error Classification.}
When evidence is insufficient, the model must determine \emph{why} the needed information was not retrieved.
It performs entity provenance checking on each search query to classify queries as \textit{Valid} or \textit{Poisoned}:
\begin{itemize}[nosep,leftmargin=1.5em]
    \item \textbf{Retriever Error}: at least one valid query (correct entities from question or prior evidence) was issued, but the retriever returned irrelevant results.
    \item \textbf{Search Error}: all queries were \emph{poisoned} by flawed reasoning (hallucinated entities, wrong assumptions, misdirected targets), so the retriever was never given a fair chance. Unlike Reasoning Error where flawed reasoning misuses sufficient evidence, here the flawed reasoning manifests in \emph{generating} invalid queries, causing the needed evidence to never be retrieved.
\end{itemize}

This integrated approach allows the diagnosis model to jointly reason about evidence sufficiency and error type, rather than relying on a separate, potentially noisy coverage predictor.

\paragraph{Shorthand notation.}
For brevity, we use the following short names throughout the paper: \textit{Format Error}, \textit{Reasoning Error}, \textit{Retriever Error}, and \textit{Search Error}.
Their full formal names (Format-Invalid Answer Error, Reasoning Logic Error, Retriever Failure, and Reasoning-Induced Missing Retrieval, respectively) are used only in the error-type definitions and the prompt templates where exact label strings are required by the model output schema.
The short form ``Search Error'' emphasizes that the failure manifests as ineffective search behavior, while the full name clarifies that the root cause lies in upstream reasoning rather than the retriever itself.

\begin{table*}[t]
\centering
\footnotesize
\setlength{\tabcolsep}{4pt}
\begin{tabular}{p{2.2cm}p{5.5cm}p{5.5cm}}
\toprule
\textbf{Error Type} & \textbf{Classification Criteria} & \textbf{What Constitutes an ``Error''} \\
\midrule
Format Error & Final answer is \emph{semantically equivalent} to GT but fails EM due to surface form, e.g., ``27'' vs.\ ``27 years old'', ``Steve Jobs'' vs.\ ``Steven Paul Jobs''. & Answer is factually correct but string-mismatched. \\
\midrule
Reasoning Error & Evidence is sufficient, but the trajectory contains a step where reasoning diverges: hallucination, contradiction, wrong entity, misinterpretation, or ignored evidence. & Sufficient evidence exists, but the model uses it incorrectly. \\
\midrule
Retriever Error & Evidence is insufficient. At least one valid query (correct entities from the question or prior evidence) is issued, but the retriever returns irrelevant or incomplete evidence. & The query targets the right information, but the retrieval fails. \\
\midrule
Search Error & Evidence is insufficient. Search queries are wrong due to flawed reasoning: hallucinated entities, wrong assumptions, misdirected targets, or excessive vagueness. & The missing evidence is caused by invalid query generation rooted in reasoning errors rather than retriever failure. \\
\bottomrule
\end{tabular}

\caption{Error type definitions and classification criteria. The diagnosis model follows a strict priority order with evidence sufficiency as the key branching point.}
\label{tab:error_criteria}
\end{table*}

\subsection{Error Criteria and Query-Level Judgement}
\label{app:diagnosis:criteria}

We define an action as erroneous only when it violates the admissible behavior of its corresponding component.
For reasoning steps, admissible behavior means that each inference is entailed by the question, prior reasoning, or retrieved evidence, and does not introduce unsupported entities or contradictions.
For search actions, admissible behavior means that the query uses correct entities from the question or previously retrieved evidence, targets a necessary missing relation or attribute, and is specific enough to give the retriever a fair chance.
For retriever behavior, admissible behavior means that a valid query should return evidence relevant to the queried entity or relation; if it fails to do so, the error is attributed to the retriever.

Table~\ref{tab:error_criteria} summarizes the criteria for each error type.
The diagnosis follows a strict priority order: Format Error is checked first, followed by the evidence sufficiency assessment, and then the appropriate sub-classification.

For all cases, the model first checks whether the final answer is semantically equivalent to the ground truth.
If so, the error is classified as \textit{Format Error}.
Otherwise, the model assesses evidence sufficiency.
If evidence is sufficient, the remaining failure is classified as \textit{Reasoning Error}, and the model localizes the earliest divergent reasoning step.
If evidence is insufficient, the model performs query-level judgement to distinguish \textit{Retriever Error} from \textit{Search Error}.

Each query is classified as either \textit{Valid} or \textit{Poisoned}.
A query is \textbf{Valid} if all entities in the query come from the user question or previously retrieved evidence, the query targets information relevant to answering the question, and is specific enough to retrieve the missing evidence.
If such a valid query fails to retrieve the needed evidence, the failure is attributed to the retriever.

A query is \textbf{Poisoned} if it is generated from flawed reasoning.
Typical poisoned queries include queries with hallucinated entities (names, dates, or facts absent from the question and all prior evidence), wrong assumptions (assuming a fact never established in the trajectory), wrong directions (targeting the wrong entity/relationship due to a prior reasoning error), or excessive vagueness.
If all search queries are poisoned, the failure is classified as \textit{Search Error}.

An important edge case is that a query may look well-formed but still target the wrong entity because of an earlier incorrect assumption.
In this case, the query is treated as poisoned, and the error is classified as \textit{Search Error} rather than \textit{Retriever Error}.
The diagnosis requires an \textit{entity provenance check}: for each key entity in a search query, the model verifies whether it originates from the user question or previously retrieved evidence (valid source) or was invented by the model in a reasoning step without evidence support (invalid source).

\subsection{Failure Localization and Repair Interface}
\label{app:diagnosis:localization}

Beyond assigning an error type, DR-RAG also localizes the earliest failure point.
This localization is essential because the repair module reuses the valid prefix before the failure point and only regenerates the affected part of the trajectory.

The diagnosis model returns \texttt{culprit\_step\_label} as the localization field for all error types.
If the error is \textit{Reasoning Error}, the culprit is the earliest reasoning step where the trajectory diverges from a valid solution path (e.g., \texttt{"[Step 3] (Reasoning)"}).
If the error is \textit{Retriever Error}, the culprit is the evidence block returned by a valid search action that fails to contain the necessary information (e.g., \texttt{"[Evidence 2] (Information)"}).
If the error is \textit{Search Error}, the culprit is the first poisoned search action---i.e., the earliest search query that is misdirected due to flawed upstream reasoning (e.g., \texttt{"[Action 2] (Search)"}).
For \textit{Format Error}, no localization is needed, and the field is set to \texttt{null}.

The final diagnosis output is denoted as $(c, k^\dagger)$, where $c$ is the assigned error type and $k^\dagger$ is the localized failure point.
This pair serves as the interface between diagnosis and repair.
The error type $c$ selects the repair operator, while the failure location $k^\dagger$ specifies the point from which local repair should be performed.


\subsection{Teacher and Student Prompts}\label{app:diagnosis:prompts}

Figure~\ref{app:prompt:unified} presents the diagnosis prompt used for both teacher annotation and student inference.
The teacher and student use the same prompt: both receive the question, trajectory, and ground-truth answer, and follow the same diagnostic logic.
The teacher's advantage lies only in its stronger reasoning capability (Qwen3.5-Plus in thinking mode), which produces more reliable diagnostic rationales for distillation.

The prompt encodes the complete diagnostic protocol as a four-phase hierarchy:
(1)~format checking,
(2)~evidence sufficiency assessment,
(3)~reasoning error localization (if evidence sufficient),
(4)~retrieval error classification via entity provenance checking (if evidence is insufficient).
The model implicitly determines evidence sufficiency during its reasoning process, eliminating the need for an external coverage signal.

A key design choice is that the ground-truth answer is provided with an explicit \textit{Assumption} statement (``The Ground Truth Answer is CORRECT. Do not question or second-guess it.'').
This eliminates the need for a separate Dataset Noise category by constraining the model to accept the ground truth as given, focusing the diagnosis entirely on identifying why the model's trajectory failed to produce the correct answer.


\section{Repair Module Details}
\label{app:repair}

This appendix provides the detailed implementation of the DR-RAG repair module, including the algorithmic procedure for each repair operator and the prompt templates used at inference time.
Given a diagnosis output $(c, k^\dagger)$---where $c$ is the error type and $k^\dagger$ is the localized failure point---the repair module selects and applies a specialized operator to correct the trajectory with minimal recomputation.

All repair operators share a common design principle: they preserve the conditionally valid prefix $y_{\mathrm{prefix}} = \{\rho^{(1)}, \ldots, \rho^{(k^\dagger-1)}\}$ and intervene only at or after the failure point.
The repair model uses the same backbone LLM as the original agentic RAG system under each experimental setting, with prompt-guided generation using vLLM.
Token budgets are enforced via truncation with a safety buffer of 50 tokens, and the maximum prompt length is capped at 30,000 tokens.

\subsection{Format-Invalid Answer Repair}
\label{app:repair:format}

Format errors occur when the model's answer is semantically correct but fails an exact match due to surface-form differences.
Since the trajectory and retrieved evidence are valid, repair is limited to rewriting the final answer.

The repair prompt (Figure~\ref{app:prompt:format_repair}) instructs the model to:
\begin{enumerate}[nosep,leftmargin=1.5em]
    \item Analyze the dialogue history (including all retrieved documents and reasoning steps) to identify the correct factual answer.
    \item Determine whether the exact answer string can be found verbatim in the retrieved documents (Priority~1: document extraction) or must be derived from reasoning (Priority~2: format enumeration).
    \item Generate a comprehensive JSON list of all valid exact-match variants (at least 10), covering: alternate surface forms, number formats, article variations, common shortenings, and symbol differences.
\end{enumerate}

The output is a JSON list of candidate strings. During evaluation, if any candidate matches the ground truth under an exact match, the repair is considered successful.

\subsection{Reasoning Logic Error Repair}
\label{app:repair:reasoning}

Reasoning errors occur when evidence is sufficient, but the model's reasoning diverges from the correct solution path.
The trajectory is truncated at the localized error index $k^\dagger$, retaining the conditionally valid prefix $y_{\mathrm{prefix}} = \{\rho^{(1)}, \ldots, \rho^{(k^\dagger-1)}\}$.
All retrieved documents $D(y)$ from the trajectory are aggregated, and the repair model re-performs reasoning from the retained prefix, grounded on $D(y)$, to generate a corrected answer (Figure~\ref{app:prompt:reasoning_repair}).
The model is instructed to base its reasoning strictly on the provided evidence and output the final answer in \texttt{<answer>} tags.

\begin{algorithm*}[t]
\small
\caption{Tool-Conditioned Local Repair}
\label{alg:repair}
\KwIn{
Failed trajectory $(q, y, a_{\mathrm{pred}})$;
diagnosis result $(c, k^\dagger)$;
retrieval service $\mathcal{R}$;
max search turns $T$.
}
\KwOut{Repaired answer $a_{\mathrm{repaired}}$.}

\BlankLine

\Switch{$c$}{
  \Case{\textit{Format Error}}{
    Prompt the model with $(q, y)$ to generate EM-valid answer variants\;
    Return the variant list as $a_{\mathrm{repaired}}$\;
  }
  \Case{\textit{Reasoning Error}}{
    Extract valid prefix $y_{\mathrm{prefix}} \leftarrow y[1:k^\dagger{-}1]$\;
    Aggregate all retrieved documents $D_{all}(y)$ from the trajectory\;
    Prompt the model with $(q, y_{\mathrm{prefix}}, D_{all}(y))$ to re-reason and generate answer\;
    Return $a_{\mathrm{repaired}}$\;
  }
  \Case{\textit{Retriever Error}}{
    Extract valid prefix $y_{\mathrm{prefix}} \leftarrow y[1:k^\dagger{-}1]$\;
    Extract queries whose retrieval results failed at and after $k^\dagger$: $\mathcal{Q}_{\mathrm{failed}}$\;
    Rewrite each $q_i \in \mathcal{Q}_{\mathrm{failed}}$ into alternative formulations $\{q_i^{(1)}, q_i^{(2)}, q_i^{(3)}\}$\;
    Re-retrieve: $D_{\mathrm{new}} \leftarrow \bigcup_j \mathcal{R}(q_i^{(j)}, \text{top-}k)$\;
    Prompt the model with $(q, y_{\mathrm{prefix}}, D_{\mathrm{new}})$ to generate answer\;
    Return $a_{\mathrm{repaired}}$\;
  }
  \Case{\textit{Search Error}}{
    Extract valid prefix $y_{\mathrm{prefix}} \leftarrow y[1:k^\dagger{-}1]$\;
    \textbf{Turn 0}: Generate search plan from $(q, y_{\mathrm{prefix}})$\;
    Initialize new history $h \leftarrow \text{plan}$\;
    \For{$t = 1$ \KwTo $T$}{
      Prompt the model with $(q, y_{\mathrm{prefix}}, h, t, T)$\;
      Parse output: \texttt{<reason>} + action\;
      \If{action is \texttt{<search>query</search>}}{
        $d \leftarrow \mathcal{R}(\text{query}, \text{top-}k)$\;
        Append reason, query, and $d$ to $h$\;
      }
      \ElseIf{action is \texttt{<answer>result</answer>}}{
        Return result as $a_{\mathrm{repaired}}$\;
      }
    }
    Force answer generation on final turn\;
    Return $a_{\mathrm{repaired}}$\;
  }
}
\end{algorithm*}

\subsection{Retriever Error Repair}
\label{app:repair:retriever}

Retriever errors occur when the model issues valid search queries, but the retriever fails to return relevant documents.
The trajectory is truncated at $k^\dagger$, retaining the valid prefix $y_{\mathrm{prefix}}$.
The repair procedure has three stages: query rewriting, re-retrieval, and answer generation.

\paragraph{Stage 1: Query Rewriting.}
The failed search queries (at $k^\dagger$ and any subsequent failed retrievals) are rewritten into multiple alternative formulations (Figure~\ref{app:prompt:query_rewrite}).
For each original query, the model generates at least 3 alternative versions using different wording, synonyms, or broader/narrower expressions, while preserving the original information needed.

\paragraph{Stage 2: Re-Retrieval.}
Each rewritten query is submitted to the retrieval service with an increased top-$k$ to maximize recall.
The retrieved documents from all rewritten queries are aggregated.

\paragraph{Stage 3: Answer Generation.}
The model generates a final answer using the original question, the retained valid prefix, and the newly retrieved evidence (Figure~\ref{app:prompt:retriever_answer}).
The model is instructed to use only the provided evidence and output the answer in \texttt{<answer>} tags.

\subsection{Search Error Repair}
\label{app:repair:search}

Search errors occur when flawed reasoning produces invalid search queries, preventing the retriever from ever receiving a fair query.
Since the root cause lies in the reasoning that generated the queries, repair requires regenerating both reasoning and retrieval from the failure point.
This is the most complex repair operator, implemented as a multi-turn planning and execution loop.

\paragraph{Stage 1: Plan Generation (Turn 0).}
The model receives the valid trajectory prefix (up to $k^\dagger$) and generates a structured search plan (Figure~\ref{app:prompt:search_plan}).
The plan decomposes the question into atomic facts, labels each as SUPPORTED (present in history) or MISSING, and specifies concrete search intents for each missing fact.
This prevents redundant searches for already-known facts.

\paragraph{Stage 2: Iterative Search and Reasoning (Turns 1--$N$).}
After planning, the model iteratively executes the plan (Figure~\ref{app:prompt:search_execute}).
Each turn produces:
\begin{enumerate}[nosep,leftmargin=1.5em]
    \item A \texttt{<reason>} block analyzing the current state and deciding the next action.
    \item Exactly one action: either \texttt{<search>query</search>} to retrieve missing information, or \texttt{<answer>result</answer>} when sufficient information is gathered.
\end{enumerate}

Search results are appended to the context for subsequent turns.
On the final turn, the model is forced to output an answer regardless of information completeness.

\subsection{Repair Algorithm Summary}
\label{app:repair:algorithm}

Algorithm~\ref{alg:repair} summarizes the complete repair procedure.
The key properties are:
(1) prefix reuse---all operators preserve the valid prefix before $k^\dagger$;
(2) minimal intervention---each operator modifies only the components required by its error type;
(3) bounded computation---the search error repair has a fixed turn budget to prevent unbounded generation.

\begin{figure*}[t]
\centering
\begin{tcolorbox}[
  enhanced,
  colback=cyan!2,
  colframe=cyan!45!black,
  arc=4pt,
  boxrule=0.6pt,
  width=\textwidth,
  title={\textbf{Diagnosis Prompt} (Teacher Annotation \& Student Inference)},
  coltitle=black,
  colbacktitle=cyan!10,
  boxed title style={colframe=cyan!45!black,colback=cyan!10,arc=4pt},
  attach boxed title to top left={xshift=6pt,yshift*=-\dimexpr\tcboxedtitleheight/2\relax},
  fontupper=\footnotesize,
  top=4pt,bottom=4pt,left=6pt,right=6pt
]

\textbf{You are an expert Fault Localization Judge for an Agentic RAG system.}
The model's answer failed Exact Match (EM=0). Your goal is to identify the root cause of the failure.
\textbf{Assumption}: The Ground Truth Answer is CORRECT. Do not question or second-guess it. Your job is to find WHY the model failed, not whether the Ground Truth is valid.

\vspace{0.3em}
\textbf{Diagnosis Logic} (Execute strictly in this order):

\textbf{Phase 1: Format-Invalid Answer Error (Priority 1 --- Check this FIRST)} ---
\textbf{Check}: Extract the model's final answer. Compare it to the Ground Truth Answer.
\textbf{IS a Format Error} (semantically same, string different): Same entity with different surface form (``Barry Bonds'' vs ``Barry Lamar Bonds''); same value with different expression (``27'' vs ``27 years old''); synonym or equivalent category (``film'' vs ``movie''); model answered correctly via parametric knowledge despite potentially incomplete evidence.
\textbf{NOT a Format Error} (different entity or factually wrong): Completely different entities; different people; different facts; model answered the wrong aspect of the question.
\textbf{Decision}: Output \textbf{``Format-Invalid Answer Error''} ONLY if the answer is the same fact in a different string form.

\textbf{Phase 2: Evidence Sufficiency Check (Key Branching Point)} ---
After ruling out Format Error, determine whether the retrieved evidence is sufficient to answer correctly:
\textbf{Evidence IS Sufficient}: [Evidence] blocks contain the necessary facts to derive the GT Answer. The model HAD the right information but failed to use it $\Rightarrow$ Go to Phase~3.
\textbf{Evidence is NOT Sufficient}: [Evidence] blocks do NOT contain the key facts needed $\Rightarrow$ Go to Phase~4.

\textbf{Phase 3: Reasoning Logic Error (Evidence Sufficient)} ---
\textbf{Condition}: Retrieved evidence is sufficient, but the model still failed.
\textbf{Criteria}: Hallucination, Contradiction, Wrong Entity/Date, Misinterpretation, Ignored Evidence.
\textbf{Decision}: $\Rightarrow$ \textbf{``Reasoning Logic Error''} + identify the culprit reasoning step.

\textbf{Phase 4: Retrieval Error Classification (Evidence NOT Sufficient)} ---
\textbf{Entity Provenance Check}: For each search query, verify entity sources:
Entity from User Question or previous Evidence $\Rightarrow$ Valid source.
Entity invented/hallucinated by model $\Rightarrow$ Invalid source.

\textbf{Phase 4a --- Retriever Failure}: Valid query (correct entities, specific) but retriever returned irrelevant results $\Rightarrow$ \textbf{``Retriever Failure''} (culprit = the failed evidence block).

\textbf{Phase 4b --- Reasoning-Induced Missing Retrieval}: Query is wrong due to hallucinated entities, wrong assumption, or misdirected targets $\Rightarrow$ \textbf{``Reasoning-Induced Missing Retrieval''} (culprit = the first poisoned search action).

\textbf{Edge case}: Well-formed query targeting wrong entity (hallucinated) $\Rightarrow$ Phase 4b, not 4a.

\vspace{0.3em}
\textbf{Input:} \texttt{User Question}: \{\textit{question}\} \quad \texttt{Ground Truth Answer}: \{\textit{gt\_answer}\} \quad \texttt{Model Trajectory} (Indexed Steps): \{\textit{formatted\_trajectory}\}

\textbf{Output} (strictly JSON): \texttt{\{"assigned\_error\_type": "...", "culprit\_step\_label": "..."\}}

\end{tcolorbox}
\caption{Diagnosis prompt used for both teacher annotation and student inference. The prompt covers all four error types with evidence sufficiency as an implicit branching point, eliminating the need for an external coverage signal.}
\label{app:prompt:unified}
\end{figure*}


\subsection{Repair Prompt Templates}
\label{app:repair:prompts}

Figures~\ref{app:prompt:format_repair}--\ref{app:prompt:search_execute} present the complete prompt templates for all repair operators.

\begin{figure*}[t]
\centering
\begin{tcolorbox}[
  enhanced,
  colback=gray!2,
  colframe=gray!50!black,
  arc=4pt,
  boxrule=0.6pt,
  width=\textwidth,
  title={\textbf{Format-Invalid Answer Repair Prompt}},
  coltitle=black,
  colbacktitle=gray!12,
  boxed title style={colframe=gray!50!black,colback=gray!12,arc=4pt},
  attach boxed title to top left={xshift=6pt,yshift*=-\dimexpr\tcboxedtitleheight/2\relax},
  fontupper=\footnotesize,
  top=4pt,bottom=4pt,left=6pt,right=6pt
]

\textbf{You are a Senior Text Rewriting and Format Correction Specialist.}
Your task is to analyze the entire dialogue history, particularly the retrieval results and reasoning, to generate a comprehensive JSON list of all possible Exact Match (EM) answers for the original question.

\vspace{0.3em}
\textbf{I. Determine the Golden Answer Format and Source}

The model's original answer is factually correct but suffers from a Format Error. Determine the correct EM answer based on:

\textbf{Priority 1: Document Extraction} --- Strictly review content inside all \texttt{<information>} tags. If the final answer can be found verbatim in retrieved documents, use that document's original expression as the primary answer.

\textbf{Priority 2: Format Enumeration} --- If not explicitly found, derive the core fact and generate all valid EM variations.

\vspace{0.3em}
\textbf{II. Generate All EM Variants}

Generate at least 10 valid variations including: alternate symbol/connector forms, number formats, common shortenings, article variations, and casing permutations.

\textbf{Requirements}: Accuracy (factually correct), Minimality (Minimal EM standard), Exhaustiveness (all common variations).

\vspace{0.3em}
\textbf{III. Output} --- Strict JSON list of strings. No reasoning or markdown outside the JSON.

\textbf{Example}: \texttt{["Original Document Phrase", "Variant 2", "Variant 3"]}

\vspace{0.3em}
{\small\textit{Input: Dialogue History (reasoning + tool calls), Original Question, Model's Incorrect Final Answer.}}

\end{tcolorbox}
\caption{Format repair prompt. The model generates multiple exact-match candidate strings from the existing trajectory without modifying reasoning or retrieval.}
\label{app:prompt:format_repair}
\end{figure*}

\begin{figure*}[t]
\centering
\begin{tcolorbox}[
  enhanced,
  colback=green!2,
  colframe=green!45!black,
  arc=4pt,
  boxrule=0.6pt,
  width=\textwidth,
  title={\textbf{Reasoning Logic Error Repair Prompt}},
  coltitle=black,
  colbacktitle=green!10,
  boxed title style={colframe=green!45!black,colback=green!10,arc=4pt},
  attach boxed title to top left={xshift=6pt,yshift*=-\dimexpr\tcboxedtitleheight/2\relax},
  fontupper=\footnotesize,
  top=4pt,bottom=4pt,left=6pt,right=6pt
]

\textbf{Question:} \{\textit{question}\}

\textbf{Valid Prefix (reasoning history before the error):}
\{\textit{valid\_prefix}\}

\textbf{Retrieved Evidence:}
\{\textit{aggregated\_documents}\}

\textbf{Instruction:}
\begin{itemize}[nosep,leftmargin=1.5em]
\item Carefully analyze the Question using the valid prefix and the retrieved evidence.
\item You may perform step-by-step reasoning to reach the answer.
\item Base your reasoning strictly on the provided evidence.
\end{itemize}

\textbf{Final Output Requirement:}
\begin{itemize}[nosep,leftmargin=1.5em]
\item Provide the final answer wrapped strictly in \texttt{<answer>}...\texttt{</answer>}.
\item If the answer can be found verbatim in the documents, use that expression.
\item The answer must be a single entity, name, or short phrase. Do NOT write complete sentences.
\end{itemize}

\end{tcolorbox}
\caption{Reasoning repair prompt. Given the valid prefix and aggregated retrieved documents, the model re-reasons to produce a corrected answer.}
\label{app:prompt:reasoning_repair}
\end{figure*}

\begin{figure*}[t]
\centering
\begin{tcolorbox}[
  enhanced,
  colback=blue!2,
  colframe=blue!45!black,
  arc=4pt,
  boxrule=0.6pt,
  width=\textwidth,
  title={\textbf{Query Rewrite Prompt} (Retriever Error Repair --- Stage 1)},
  coltitle=black,
  colbacktitle=blue!10,
  boxed title style={colframe=blue!45!black,colback=blue!10,arc=4pt},
  attach boxed title to top left={xshift=6pt,yshift*=-\dimexpr\tcboxedtitleheight/2\relax},
  fontupper=\footnotesize,
  top=4pt,bottom=4pt,left=6pt,right=6pt
]

\textbf{User Question:} \{\textit{question}\}

The agent previously issued the following failed or ineffective search queries:

\{\textit{failed\_queries}\}

\textbf{Your Task:} For EACH original query:
\begin{enumerate}[nosep,leftmargin=1.5em]
\item Analyze its underlying information need.
\item Rewrite it into AT LEAST 3 alternative search queries.
\item The rewritten queries should: preserve the original intent; use different wording, synonyms, or broader/narrower expressions; be suitable for dense or hybrid retrieval systems.
\end{enumerate}

\textbf{Important Rules:}
\begin{itemize}[nosep,leftmargin=1.5em]
\item Do NOT merge different original queries. Handle each independently.
\item Each original query must have its own rewritten versions.
\end{itemize}

\textbf{Output Format:} For each original query, output rewritten versions in \texttt{<search>} tags.

{\footnotesize
\texttt{Original Query: xxx}\\
\texttt{<search>rewrite 1</search>}\\
\texttt{<search>rewrite 2</search>}\\
\texttt{<search>rewrite 3</search>}
}

\end{tcolorbox}
\caption{Query rewrite prompt. Each failed query is independently rewritten into multiple alternative formulations to improve retrieval recall.}
\label{app:prompt:query_rewrite}
\end{figure*}

\begin{figure*}[t]
\centering
\begin{tcolorbox}[
  enhanced,
  colback=blue!2,
  colframe=blue!45!black,
  arc=4pt,
  boxrule=0.6pt,
  width=\textwidth,
  title={\textbf{Answer Generation Prompt} (Retriever Error Repair --- Stage 3)},
  coltitle=black,
  colbacktitle=blue!10,
  boxed title style={colframe=blue!45!black,colback=blue!10,arc=4pt},
  attach boxed title to top left={xshift=6pt,yshift*=-\dimexpr\tcboxedtitleheight/2\relax},
  fontupper=\footnotesize,
  top=4pt,bottom=4pt,left=6pt,right=6pt
]

\textbf{Question:} \{\textit{question}\}

\textbf{--- Retrieved Evidence ---}

The following documents are retrieved from the knowledge base. They may contain overlapping or complementary information.

\{\textit{docs\_block}\}

\textbf{--- Instruction ---}

Using ONLY the information in the retrieved evidence above, answer the question as accurately as possible.
Wrap your final answer strictly inside \texttt{<answer>} and \texttt{</answer>}.
If the answer can be found verbatim in the documents, use that expression.
The answer must be a single entity, name, or short phrase. Do NOT write complete sentences.

\end{tcolorbox}
\caption{Answer generation prompt after re-retrieval. The model synthesizes the newly retrieved evidence to produce a corrected answer.}
\label{app:prompt:retriever_answer}
\end{figure*}

\begin{figure*}[t]
\centering
\begin{tcolorbox}[
  enhanced,
  colback=orange!2,
  colframe=orange!45!black,
  arc=4pt,
  boxrule=0.6pt,
  width=\textwidth,
  title={\textbf{Search Plan Prompt} (Search Error Repair --- Turn 0)},
  coltitle=black,
  colbacktitle=orange!10,
  boxed title style={colframe=orange!45!black,colback=orange!10,arc=4pt},
  attach boxed title to top left={xshift=6pt,yshift*=-\dimexpr\tcboxedtitleheight/2\relax},
  fontupper=\footnotesize,
  top=4pt,bottom=4pt,left=6pt,right=6pt
]

\textbf{--- TASK ---}

You are repairing a failed reasoning trajectory. Your task is to RE-PLAN a correct solution path.

\textbf{[History Start]}\\
\{\textit{old\_history (valid prefix)}\}\\
\textbf{[History End]}

\textbf{Question:} \{\textit{question}\}

\textbf{--- DEFINITIONS ---}
\begin{itemize}[nosep,leftmargin=1.5em]
\item The history may already contain VALID facts from previous searches.
\item Any information present in the history CAN be reused directly.
\item You MUST NOT plan new searches for facts already supported by the history.
\end{itemize}

\textbf{--- INSTRUCTIONS ---}

Based strictly on the history and question, construct a SEARCH-DRIVEN execution plan. The goal is to obtain ONLY the missing knowledge needed to answer the question.

Your plan MUST be enclosed in \texttt{<plan>} and \texttt{</plan>} tags. In the plan:
\begin{enumerate}[nosep,leftmargin=1.5em]
\item List all atomic facts REQUIRED to answer the question.
\item For EACH fact, label as SUPPORTED or MISSING.
\item For EACH MISSING fact, specify ONE concrete \texttt{<search>} intent.
\item Define the search order.
\end{enumerate}

\textbf{Constraints}: Do NOT include searches for SUPPORTED facts. Do NOT perform reasoning, searching, or answering. ONLY output the \texttt{<plan>}.

\end{tcolorbox}
\caption{Search plan prompt (Turn~0). The model generates a structured plan identifying which facts are missing and what searches are needed.}
\label{app:prompt:search_plan}
\end{figure*}

\begin{figure*}[t]
\centering
\begin{tcolorbox}[
  enhanced,
  colback=orange!2,
  colframe=orange!45!black,
  arc=4pt,
  boxrule=0.6pt,
  width=\textwidth,
  title={\textbf{Search Execution Prompt} (Search Error Repair --- Turns 1--$N$)},
  coltitle=black,
  colbacktitle=orange!10,
  boxed title style={colframe=orange!45!black,colback=orange!10,arc=4pt},
  attach boxed title to top left={xshift=6pt,yshift*=-\dimexpr\tcboxedtitleheight/2\relax},
  fontupper=\footnotesize,
  top=4pt,bottom=4pt,left=6pt,right=6pt
]

\textbf{\#\#\# System Role}

You are a Correction Agent. Your goal is to fix a reasoning path and guide it to the correct answer.

\textbf{\#\#\# Output Format Rules}
\begin{enumerate}[nosep,leftmargin=1.5em]
\item \texttt{<reason>}: Analyze the next step. Strictly plain text.
\item \textbf{Action}: Immediately follow reasoning with EXACTLY ONE block:
  \begin{itemize}[nosep,leftmargin=1em]
  \item \texttt{<search>query</search>}: Retrieve missing external info.
  \item \texttt{<answer>result</answer>}: Provide the final answer (single entity/short phrase).
  \end{itemize}
\end{enumerate}

\textbf{\#\#\# Input Data}

\textbf{Question:} \{\textit{question}\}

\textbf{Context History:} \{\textit{old\_history}\}

\textbf{Current Corrective Reasoning:} \{\textit{new\_history}\}

\textbf{\#\#\# Instructions}
\begin{enumerate}[nosep,leftmargin=1.5em]
\item Reason: Output reasoning for the correction in a \texttt{<reason>} tag.
\item Act: Take the first correct action (\texttt{<search>} or \texttt{<answer>}).
\end{enumerate}

\{\textit{turn\_constraint}\}

\vspace{0.3em}
{\footnotesize \textit{turn\_constraint} = ``This is the LAST TURN. You MUST output \texttt{<answer>} immediately.'' on the final turn; otherwise ``You may use \texttt{<search>} if needed, or \texttt{<answer>} if sufficient information is gathered.''}

\end{tcolorbox}
\caption{Search execution prompt (Turns 1--$N$). The model iteratively searches for missing facts and reasons toward the final answer, with a forced-answer constraint on the last turn.}
\label{app:prompt:search_execute}
\end{figure*}


%


\end{document}